\documentclass[
reprint,
bibnotes, longbibliography
amsmath,amssymb,
aps,pra,showkeys,
floatfix,
]{revtex4-2}

\usepackage{graphicx,xcolor}
\usepackage{dcolumn}
\usepackage{bm}
\usepackage{orcidlink}
\usepackage{amsmath,amssymb}

\DeclareMathOperator{\erf}{erf}
\DeclareMathOperator{\adj}{adj}

\DeclareMathOperator{\arcsinh}{arcsinh}
\DeclareMathOperator{\HeunC}{HeunC}    

\begin{document}
\preprint{APS/123-QED}

\title{Geometric effects on the electronic structure of curved nanotubes and curved graphene: the case of the helix, catenary, helicoid, and catenoid}
\author{J.D.M. de Lima \orcidlink{0000-0002-4296-6001}}
\author{E. Gomes \orcidlink{0000-0002-3812-9331}}
\author{F.F. da Silva Filho \orcidlink{0000-0003-3832-787X}}
\author{F. Moraes \orcidlink{0000-0001-7045-054X}}
 \email{fernando.jsmoraes$@$ufrpe.br}
\affiliation{Departamento de Física, Universidade Federal Rural de Pernambuco, Recife, PE, 52171-900, Brazil}
\author{R. Teixeira \orcidlink{0000-0003-0314-8142}}
\affiliation{Departamento de Matemática, Universidade Federal Rural de Pernambuco, Recife, PE, 52171-900, Brazil}

\date{\today}

\begin{abstract}
    Since electrons in a ballistic regime perceive a carbon nanotube or a graphene layer structure as a continuous medium, we can use the study of the quantum dynamics of one electron constrained to a curve or surface to obtain a qualitative description of the conduction electrons' behavior. The confinement process of a quantum particle to a curve or surface leads us, in the so-called ``confining potential formalism'' (CPF), to a geometry-induced potential (GIP) in the effective Schrödinger equation. With these considerations, this work aims to study in detail the consequences of constraining a quantum particle to a helix, catenary, helicoid, or catenoid, exploring the relations between these curves and surfaces using differential geometry. Initially, we use the variational method to estimate the energy of the particle in its ground state, and thus, we obtain better approximations with the use of the confluent Heun function through numerical calculations. Thus, we conclude that a quantum particle constrained to an infinite helix has its angular momentum quantized due to the geometry of the curve, while in the cases of the catenary, helicoid, and catenoid the particle can be found either in a single bound state or in  excited states which constitute a continuous energy band. {Additionally, we propose measurements of physical observables capable of discriminating the topologies of the studied surfaces, in the context of topological metrology.  }
\end{abstract}
\maketitle

\section{Introduction}
Due to the great advance of materials science, it became possible to synthesize increasingly smaller carbon-based structures. Along with the discovery of graphene, the first truly two-dimensional system, several theoretical techniques have been proposed to describe the quantum dynamics of its charge carriers. Many other carbon nanostructures have also  been synthesized with different shapes, such as fullerenes, nanotubes, nanocones, and structures with non-trivial geometries \cite{terrones2003curved}. In this sense, it is necessary to understand the effects of the geometry on the quantum dynamics of a particle in such curved structures. 

One of the first formal approaches addressed to solve this problem was proposed by H. Jensen and H. Koppe in 1971, which studied quantum particles constrained to move between two parallel surfaces in the limit where the inter-surface distance goes to zero \cite{jensen1971quantum}. In 1981, R.C.T. da Costa considered that the constraint procedure can be realized through a smooth potential well \cite{da1981quantum,da1982constraints}. In both cases, it was showed that it leads to a geometry-induced potential (GIP) which acts upon the dynamics. If the particle is constrained to a space curve, the GIP  depends on its  curvature. On the other hand, when dealing with curved surfaces the GIP dependence is on its mean and Gaussian curvatures. This is known \cite{schuster2003quantum} as ``confining potential formalism'' (CPF). Experimental verification of geometric effects due to confinement were verified first in photonic topological crystals \cite{szameit2010geometric} and in a peanut-shaped $C_{60}$ polymer \cite{onoe2012observation}.

As the electronic transport in graphene is mesoscopically ballistic, such electrons perceive the structure as a continuous medium \cite{zhang2014defects}. This implies that the study of the quantum dynamics of an electron confined to a surface (or a curve) leads to a qualitative description of the behavior of electrons in the ballistic regime in a graphene layer (or in a carbon nanotube) with topological defects, such that the geometry has an important role in the dynamics \cite{santos2016geometric}.

{The study of curved structures can be essential for the understanding of their properties and possible applications, {especially for electronic devices. For example, Ref. \cite{joglekar2009curvature} suggests that p-n junctions can be induced by curvature in a graphene bilayer, whose geometry is related to a catenoid \cite{silva2020electronic}.} In Ref. \cite{atanasov2009geometry} it was shown that a helicoidal ribbon can induce, due to its curvature, a charge separation which results in a electrical field crossing the molecule.}
{
Furthermore, the catenoid can be considered in the study of wormholes \cite{dandoloff2004quantum,azevedo2021optical} and as an approximation to the geometries of ion-channel membrane proteins \cite{gupta2012negative,gupta2011geometrical}. Similarly, the helix and the helicoid can also be considered as approximations to the geometries of $\beta$-sheet proteins, DNA molecules \cite{atanasov2009geometry,gupta2012negative,gupta2018role}, and helicenes \cite{gingras2013one}, for example.}

The description  of regular curves and surfaces is very well established by differential geometry. Two of the simplest regular curves that we can study in this context are the helix and the catenary. The surface formed by continuously varying the radius of the helix is called helicoid, while the surface of revolution associated with the catenary is called catenoid. As the helicoid and the catenoid are locally isometric and both of them are minimal surfaces, their properties have been reported in detail by several authors  \cite{stokerdifferential,struik1988lectures,Manfredo}.

The confinement of a quantum particle to a helicoid or catenoid with the CPF has been the subject of some studies. V. Atanasov, R. Dandoloff, and A. Saxena applied the CPF to the helicoid in a work published in 2009, where they showed that  the constrained particle is subject to an effective radial potential that can be attractive or repulsive, depending on the angular momentum \cite{atanasov2009geometry}. For the catenoid, the CPF was used in 2010 by R. Dandoloff, A. Saxena, and B. Jensen to understand the GIP in a two-dimensional section of a wormhole \cite{dandoloff2010geometry}. Recently, in 2020, J.E.G. Silva et al. used the electron confinement to the catenoid to study the electronic properties of bilayer graphene connected by a bridge \cite{silva2020electronic}. Also in 2020, M.C.R. Ribeiro Jr. et al. studied the confinement problem on the helicoid considering the presence of an harmonic oscillator potential and anisotropic mass \cite{ribeiro2020quantum}.

With the above motivations, we study the Schr\"{o}dinger equation for a quantum particle constrained to a helix, catenary, helicoid, or catenoid, using the CPF. We intend here to understand the particularities of the confinement problem in each one of them. In section \ref{geodiff} we present a brief review of the differential geometry of curves and surfaces, while the CPF is described in section \ref{da-costa-approach}. In section \ref{Schrodinger-solutions} we solve the problem for the helix, and we introduce the problems for the catenary, helicoid, and catenoid (solving the angular part for the surfaces). In section \ref{radial-part} we estimate the ground state energy using the variational method for the cases of the catenary, helicoid, and catenoid, besides we present solutions in terms of the confluent Heun functions. In section \ref{sec:ground-state} we discuss the solutions for the ground state. Finally, {topological aspects are discussed in section \ref{topological-aspects} and}  our conclusions and perspectives for future works are presented in Section \ref{conclusions}.

\section{Overview of differential geometry}
\label{geodiff}
In this section we do a brief presentation of some concepts of differential geometry that are necessary for the understanding of this text. More details can be found in Ref. \cite{Manfredo}. The procedure that we will follow in this section is entirely focused on developing a mathematical support to be used in the discussion about the CPF.

\subsection{Curves}
Let $\boldsymbol{\alpha}$ be a regular curve in $\mathbb{R}^{3}$ whose parametrization is given by
\begin{equation}
    \boldsymbol{\alpha}(t) = \left(x(t), y(t), z(t) \right),
    \label{C-standard}
\end{equation}
where $t$ is a parameter in an open interval $T \subset \mathbb{R}$. Thus, the arc length of $\boldsymbol{\alpha}$, from some arbitrary $t_{0} \in T$, is defined by
\begin{equation}
    s(t) = \int_{t_{0}}^{t} \left|\boldsymbol{\alpha}'(\tau) \right| d\tau.
    \label{arc-length}
\end{equation}
When it is possible to obtain $s(t)$ and its inverse, $ s^{-1}(t)$, we can find an arc length parametrization by making $\boldsymbol{\gamma}(s) = \boldsymbol{\alpha} \circ s^{-1}$, which is convenient since $|\boldsymbol{\gamma}'(s)| = 1, \, \forall \,  s$. In this parametrization, the curvature of the curve is given by
\begin{eqnarray}
    \kappa(s) = \left|\boldsymbol{\gamma}''(s)\right|,
    \label{C-curvature}
\end{eqnarray}
where $\kappa(s)$ is always positive \cite{Manfredo,stokerdifferential}. 

\begin{figure}[htp]
    \centering
    \includegraphics[width=0.6\columnwidth]{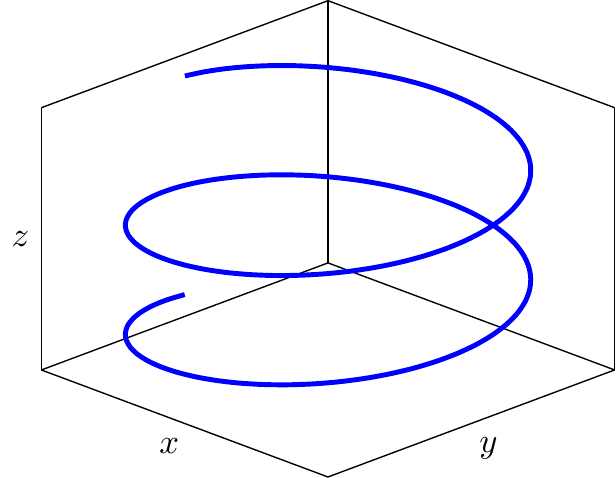}
    \caption{Three-dimensional representation of the helix. }
    \label{fig:helix}
\end{figure}

The helix (see Fig. \ref{fig:helix}) with radius $b > 0$ and pitch (vertical distance between each loop) given by $2\pi |a|$, for $a \in \mathbb{R}$, is a regular space curve parametrized by
\begin{equation}
    \boldsymbol{\alpha}(\phi) = \left(b\cos(\phi), b\sin(\phi), a\phi \right),
    \label{helix-standard}
\end{equation}
where $\phi \in \mathbb{R}$ is the curve parameter. Consequently, its arc length is $s = \sqrt{a^{2} + b^{2}} \phi$ and, after obtaining $\boldsymbol{\gamma}(s)$, we get from equation \eqref{C-curvature} the constant curvature
\begin{equation}
    \kappa(s) = \dfrac{b}{a^{2} + b^{2}}.
    \label{helix-curvature}
\end{equation}

\begin{figure}[htp]
    \centering
    \includegraphics[width=0.6\columnwidth]{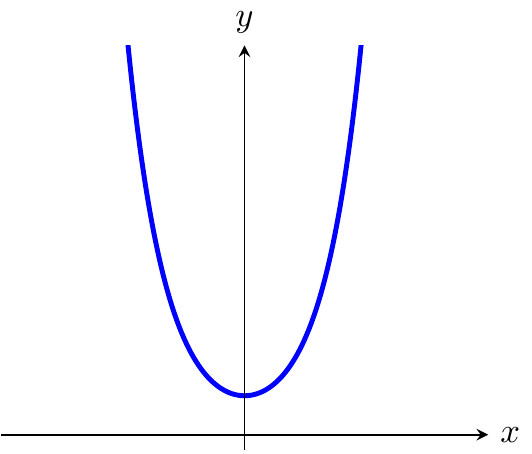}
    \caption{Two-dimensional representation of a catenary.}
    \label{fig:catenary}
\end{figure}

The catenary (see Fig. \ref{fig:catenary}) is a regular plane curve parametrized by
\begin{equation}
    \boldsymbol{\alpha}(v) = (v, a\cosh(v/a)),
    \label{catenary-standard}
\end{equation}
for $v \in \mathbb{R}$. Here, the constant $a > 0$ is associated with the catenary opening. From Eq. \eqref{arc-length} we get $s = a\sinh(v/a)$, such that \eqref{catenary-standard} gives us
\begin{equation}
    \boldsymbol{\gamma}(s) = \left(a \arcsinh(s/a), \sqrt{a^{2} + s^{2}}\right),
\end{equation}
and 
\begin{equation}
   \kappa(s) = \dfrac{a}{a^{2} + s^{2}}.
   \label{catenary-curvature}
\end{equation}

\subsection{Surfaces}
Let $S$ be a regular surface embedded in $\mathbb{R}^{3}$ whose parametrization is
\begin{equation}
    \mathbf{r}(q_{1}, q_{2}) = \left(x(q_{1}, q_{2}), y(q_{1}, q_{2}), z(q_{1}, q_{2}) \right),
\end{equation}
where $q_{1}$ and $q_{2}$ are parameters in an open set $U \subset \mathbb{R}^{2}$. According to Ref. \cite{Manfredo}, the regularity condition implies that $\mathbf{r}_{1} = \dfrac{\partial \mathbf{r}}{\partial q_{1}}$ and $\mathbf{r}_{2} = \dfrac{\partial \mathbf{r}}{\partial q_{2}}$ are linearly independent, \textit{i.e.}, $\mathbf{r}_{1} \times \mathbf{r}_{2} \neq 0$ at all points of $S$, which allows us to define the unit vector normal to the surface, $\mathbf{N} = \dfrac{\mathbf{r}_{1} \times \mathbf{r}_{2}}{|\mathbf{r}_{1} \times \mathbf{r}_{2}|}$.

The metric tensor of the surface $S$ is given by 
\begin{equation}
    \mathbf{g} = 
    \begin{pmatrix}
        g_{11} & g_{12} \\
        g_{21} & g_{22}
    \end{pmatrix},
    \label{metric-tensor}
\end{equation}
where $g_{ij} = \mathbf{r}_{i} \cdot \mathbf{r}_{j}$ (for $i, j = 1, 2$) are the coefficients of the first fundamental form. Thus, denoting by $g^{ij}$ the elements of the inverse matrix $\mathbf{g}^{-1}$, we get
\begin{equation}
       \mathbf{g}^{-1} =
       \begin{pmatrix}
            g^{11} & g^{12} \\
            g^{21} & g^{22}
        \end{pmatrix}
        = \dfrac{\adj(\mathbf{g})}{\det(\mathbf{g})}
        = \dfrac{1}{g} 
        \begin{pmatrix}
            g_{22} & - g_{12} \\
           -  g_{21} & g_{11}
        \end{pmatrix},
        \label{metric-tensor-inverse}
\end{equation}
for $g = \det(\mathbf{g})$, the determinant of the metric tensor. Further, for $\mathbf{r}_{ij} = \dfrac{\partial^{2} \mathbf{r}}{\partial q_{i} \partial q_{j}}$,  the coefficients of the second fundamental form are defined as $h_{ij} = \mathbf{N} \cdot \mathbf{r}_{ij}$. Thus, the mean and Gaussian curvatures are, respectively,
\begin{align}
    M & = \dfrac{1}{2}\dfrac{g_{11} h_{22} - 2 g_{12} h_{12} + g_{22} h_{11}}{g_{11} g_{22} - g_{12}^{2}} , \label{mean-curvature} \\
    K & = \dfrac{h_{11} h_{22} - h_{12}^{2}}{g_{11} g_{22} - g_{12}^{2}} . \label{gaussian-curvature}
\end{align}

\begin{figure}[htp!]
   \centering
    \includegraphics[width=0.65\columnwidth]{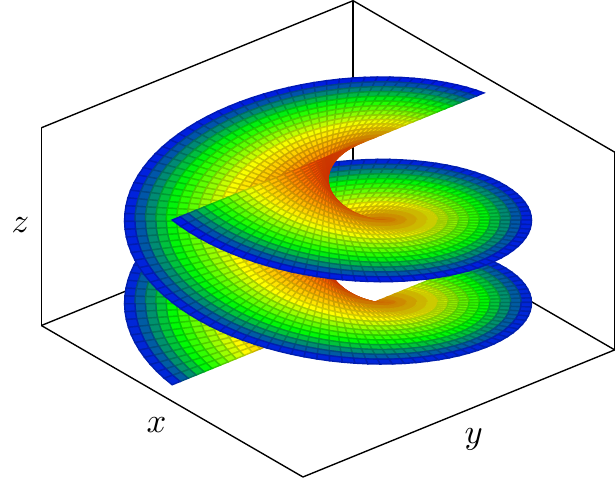}
    \caption{Three-dimensional graph of the helicoid  with $0 < \phi < 2\pi$.}
    \label{fig:helicoid}
\end{figure}

For the helicoid,  we consider $q_{1} = \phi \in \mathbb{R}$ and $q_{2} = u \in \mathbb{R}$, which are coordinates on $S$. The parametrization of the helicoid (see Fig. \ref{fig:helicoid}) is given by
\begin{equation}
    \mathbf{r}(\phi, u) = \left(u\cos(\phi), u\sin(\phi), a\phi \right),
    \label{helicoid-param}
\end{equation}
where $a$ is a real constant. As we can see, for each fixed $u$ we get a different helix. From \eqref{helicoid-param}, the coefficients of the first fundamental form are $g_{11} = a^{2} + u^{2}$, $g_{12} = g_{21} = 0$, and $g_{22} = 1${, which give us the metric
\begin{equation}
    ds^{2} = (a^{2} + u^{2})d\phi^{2} + du^{2}.
    \label{metric-helic}
\end{equation}
} In the same way, the coefficients of the second fundamental form are $h_{11} = 0$, $h_{12} = h_{21} = \dfrac{a}{\sqrt{a^{2} + u^{2}}}$, and $h_{22} = 0$. Thus, from Eqs. \eqref{mean-curvature} and \eqref{gaussian-curvature}, we get
\begin{equation}
    M = 0 \quad \text{and} \quad K = - \dfrac{a^{2}}{(a^{2} + u^{2})^{2}}.
    \label{gauss-mean-helicoid-catenoid}
\end{equation}

\begin{figure}[htp]
    \centering
    \includegraphics[width=0.65\columnwidth]{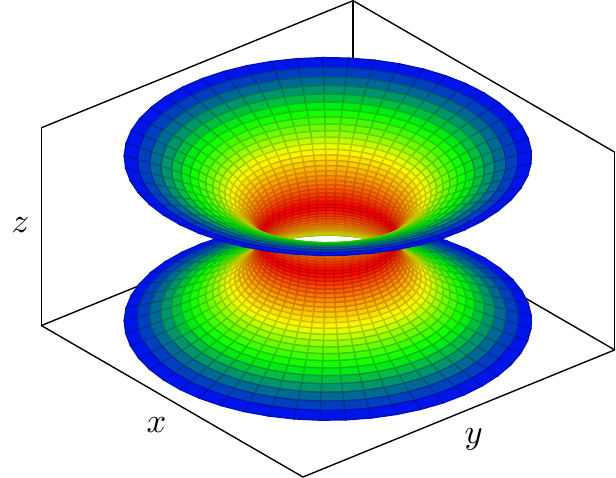}
    \caption{Three-dimensional graph of the catenoid. The color  indicates the corresponding regions of the helicoid in Fig. \ref{fig:helicoid}.}
    \label{fig:catenoid}
\end{figure}

The catenoid (see Fig. \ref{fig:catenoid}) is a surface generated by the revolution of a catenary. Making $q_{1} = \phi \in (0, 2\pi)$ and $q_{2} = u \in \mathbb{R}$, a parametrization for the catenoid is given by
\begin{equation}
    \begin{aligned}
        \mathbf{r}(\phi, u) = \Big(\sqrt{a^{2} + u^{2}} \cos(\phi), \sqrt{a^{2} + u^{2}} \sin(\phi), &\\
        a \arcsinh(u/a & )\Big),
        \label{catenoid-parametrization}
    \end{aligned}
\end{equation}
where  $a$, the ``throat radius'', is a positive constant. Thus, from \eqref{catenoid-parametrization}, we get $g_{11} = a^{2} + u^{2}$, $g_{12} = g_{21} = 0$, and $g_{22} = 1$, {which give us the same metric in Eq. \eqref{metric-helic}. The} coefficients of the second fundamental form are $h_{11} = - a$, $h_{12} = h_{21} = 0$, and $h_{22} = \dfrac{a}{a^{2} + u^{2}}$. Since the helicoid and the catenoid have the same coefficients of the first fundamental form in these parametrizations, they are locally isometric. In other words, it is possible to deform a catenoid onto a helicoid with $ 0 < \phi < 2\pi$ while maintaining the same metric. It follows from Gauss's Theorema Egregium that they must have the same Gaussian curvature. In fact, from Eqs. \eqref{mean-curvature} and \eqref{gaussian-curvature}, we get for the catenoid the same mean and Gaussian curvatures of the helicoid, which are given by Eqs. \eqref{gauss-mean-helicoid-catenoid} (cf. \cite{Manfredo,stokerdifferential}). The regions of same color in Figs. \ref{fig:helicoid} and \ref{fig:catenoid} indicate corresponding regions on the respective surfaces. In other words, the colored helicoid of Fig. \ref{fig:helicoid} becomes the catenoid of Fig. \ref{fig:catenoid} under a suitable deformation.

\section{Confining potential formalism}
\label{da-costa-approach}
In order to maintain a particle constrained to a surface $S$, it is proposed that the particle is  under the action of  forces orthogonal to $S$ in all its points, such that the ``spreading'' of the wave function in the transverse direction is ``squeezed'' \cite{jensen1971quantum,da1981quantum,da1982constraints}. Therefore, the proposal consists of considering a potential that is always constant on $S$ but increases sharply in the  normal direction, in such a way that the confinement occurs at the limit where an infinite potential keeps the particle bound to $S$ \cite{bastos2016quantum,da2017quantum}. An analogous procedure is done for curves. Thus, a GIP  appears in the effective Schr\"{o}dinger equation of a particle constrained to a curve or to a surface as a consequence of the confinement.

\subsection{Particle constrained to a curve}
Let $\boldsymbol{\gamma}$ be a regular curve (embedded in ${\mathbb{R}}^{3}$) which is parametrized by its arc length, $s$, and has curvature $\kappa = \kappa(s)$. As  shown in Refs. \cite{jensen1971quantum,da1981quantum}, a quantum particle of mass $m$, constrained to $\boldsymbol{\gamma}$, is subject to a GIP
 \begin{eqnarray}
     V_{C}(s) = -\dfrac{\hbar^{2}}{8m}\kappa^{2}.
     \label{curve-potential}
 \end{eqnarray}
  From Eq. \eqref{curve-potential} we get the time-independent Schr\"{o}dinger equation
 \begin{equation}
     -\dfrac{\hbar^{2}}{2m} \dfrac{d^{2}\psi}{ds^{2}} - \dfrac{\hbar^{2}}{8m}\kappa^{2}\psi = E\psi,
     \label{curve-schrodinger}
 \end{equation}
where $\psi = \psi(s)$ is the wave function and $E$ is the energy of the particle. The attractive potential indicates that the particle may be found in bound states  due to  geometric effects.

\subsection{Particle constrained to a surface}
Let $S$ be a regular surface embedded in ${\mathbb{R}}^{3}$ with parameters $q_{1}$ and $q_{2}$. Following the CPF, a quantum particle of mass $m$, constrained to $S$, is subjected to a GIP
\begin{equation}
    V_{S}(q_{1}, q_{2}) = - \dfrac{\hbar^{2}}{2m}\left(M^{2} - K\right),
    \label{surface-potential}
\end{equation}
where $M$ and $K$ are defined in \eqref{mean-curvature} and \eqref{gaussian-curvature}, respectively \cite{bastos2016quantum,da1982constraints,da1981quantum}. Thus, from \eqref{surface-potential} we arrive at the time-independent Schr\"{o}dinger equation
\begin{equation}
    - \dfrac{\hbar^{2}}{2m} \Delta_{g} \chi - \dfrac{\hbar^{2}}{2m}\left(M^{2} - K\right) \chi = E\chi,
    \label{schrodinger-surface}
\end{equation}
where $\chi = \chi(q_{1}, q_{2})$ is the wave function and $\Delta_{g}$ is the Laplace-Beltrami operator which is given by \cite{bernard2013notes}
\begin{equation}
    \Delta_{g}\chi = \sum_{i,j = 1}^{2} \dfrac{1}{\sqrt{g}}\dfrac{\partial}{\partial q_{i}} \left(\sqrt{g} g^{ij} \dfrac{\partial \chi}{\partial q_{j}} \right). \label{LBop}
\end{equation}
Isometric surfaces have the same Gaussian curvature but their respective mean curvatures differ \cite{da1981quantum,Manfredo}, except in the case of minimal surfaces, which have $M=0$. This is the case of the object of our study in this article, the helicoid and the catenoid,  which have the same GIPs, given by Eq. \eqref{surface-potential}.

\section{Schr\"{o}dinger equations}
\label{Schrodinger-solutions}
 In this Section, we apply Eq. \eqref{curve-schrodinger} to the helix and the catenary, and Eq. \eqref{schrodinger-surface} to the helicoid and the catenoid. In addition, we  introduce the respective boundary conditions for the wave function in both cases and  discuss their implications.

\subsection{Helix}
\label{subsec:helix-results}
From Eqs. \eqref{helix-curvature} and \eqref{curve-potential}, we get the GIP that acts on a particle constrained  to a helix,
\begin{equation}
    V_{C} = -\dfrac{\hbar^{2}}{8m}\dfrac{b^{2}}{(a^{2} + b^{2})^{2}},
    \label{helix-potential}
\end{equation}
which is constant. Thus, recalling that $s = \sqrt{a^{2} + b^{2}}\phi$ and using Eqs. \eqref{curve-schrodinger} and \eqref{helix-potential}, we  write
\begin{equation}
    \dfrac{d^{2}\psi}{d\phi^{2}} + l^{2}\psi = 0, 
    \label{helix-ode}
\end{equation}
where 
\begin{equation}
    l^{2} = \dfrac{b^{2}}{4(a^{2} + b^{2})} + \dfrac{2m(a^{2} + b^{2})}{\hbar^{2}}E . \label{Ehelixinf}
\end{equation}

Now, let us discuss the boundary conditions for Eq. \eqref{helix-ode} when we have an infinite helix ($\phi \in \mathbb{R}$). Let $p$ be an arbitrary point on the helix and $B(p)  \subset \mathbb{R}^{3}$ be a ball of radius $\varepsilon_{B}$ centered on $p$. Within that ball, the particle perceives the surrounding environment in the same way at any point of the helix, because $B$ is flat, and the helix has constant curvature and torsion. Thus, the wave function is distributed in the same way in the surrounding of $\boldsymbol{\alpha}(\phi)$ and $\boldsymbol{\alpha}(\phi + 2\pi)$. Therefore, we have the appropriate boundary condition $\psi(\phi) = \psi(\phi + 2\pi)$. This immediately leads to the quantization of the angular momentum, $\hat{L}_{\phi} = - i\hbar \dfrac{d}{d\phi}$. In other words, if $\psi$ is a eigenfunction of $\hat{L}_{\phi}$, we have the eigenvalues $l \hbar$, for $l = 0, \pm 1, \pm 2, \dots$, such that Eq. \eqref{helix-ode} is retrieved in terms of $\hat{L}_{\phi}^2$. The quantized energy is obtained from Eq. \eqref{Ehelixinf}:
\begin{equation}
    E_{l} = \dfrac{l^{2} \hbar^{2} }{2m(a^{2} + b^{2})} - \frac{\hbar^{2} b^2}{8m (a^{2} + b^{2})^2 } . \label{Einfinite}
\end{equation}
The corresponding propagating modes are degenerate, since they can move either ``up'' or ``down'' on the helix, depending on the sign of $l$. The delta-function normalized wave functions are given by
\begin{equation}
    \psi_{l}(\phi) = \frac{1}{\sqrt{2\pi}} e^{il\phi}.
\end{equation}

For a finite helix with $0 < \phi < \phi_{0}$, for some fixed $\phi_{0}$, we have the boundary conditions $\psi(0) = \psi(\phi_{0}) = 0$, such that we get $l = \dfrac{j\pi}{\phi_{0}}$, therefore, the solutions of Eq. \eqref{helix-ode} are
\begin{equation}
    \psi_{j}(\phi) = \sqrt{\dfrac{2}{\phi_{0}}} \sin\left(\dfrac{j\pi}{\phi_{0}}\phi\right),
    \label{helix-wavef}
\end{equation}
where $j = 1, 2, 3, \dots$ is a quantum number. Thus, the possible energies are given by
\begin{equation}
    E_{j} = \dfrac{ j^{2} \pi^{2} \hbar^{2}}{2m(a^{2} + b^{2}){\phi_{0}}^{2}} - \dfrac{\hbar^{2}b^{2}}{8m(a^{2} + b^{2})^{2}}.
    \label{helix-energy}
\end{equation}
Note that both Eqs. \eqref{Einfinite} and \eqref{helix-energy}   are similar to those obtained for a particle in an infinite potential well with the geometric contribution from the GIP \eqref{helix-potential}.

\subsection{Catenary}
\begin{figure}[htp]
    \centering
    \vspace{0.5cm}
   \includegraphics[width=0.9\columnwidth]{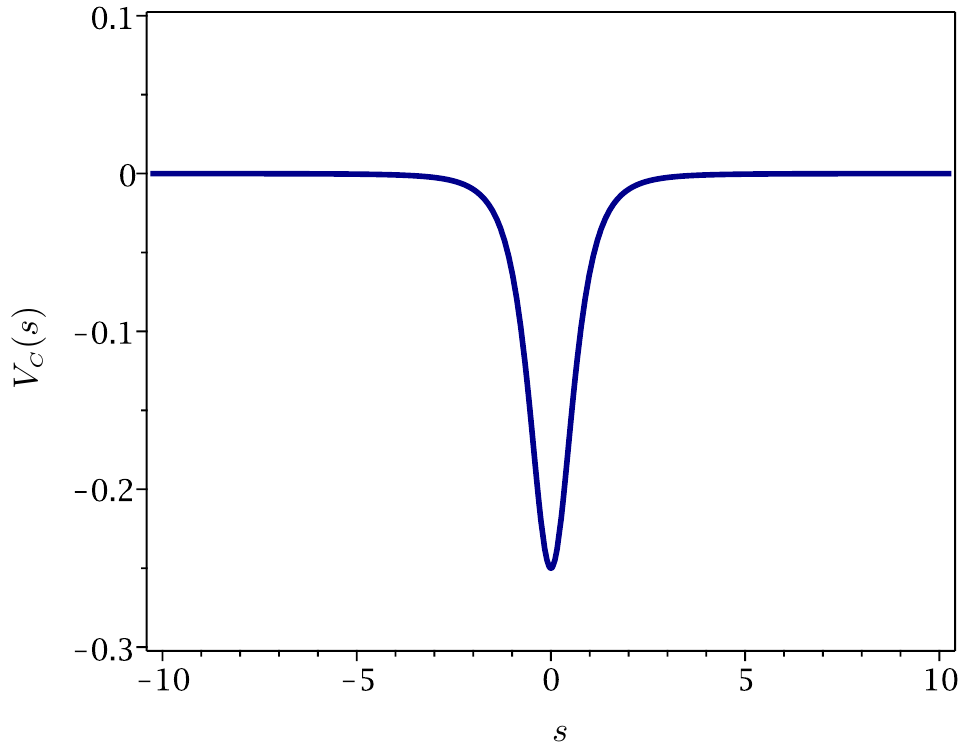}
    \caption{GIP for a  particle constrained to a catenary considering $\hbar = 2m = a = 1$.}
    \label{fig:catenarypot}
\end{figure}

From Eqs. \eqref{catenary-curvature} and \eqref{curve-potential} we get
\begin{equation}
    V_{C}(s) = - \dfrac{\hbar^{2}}{8m} \dfrac{a^{2}}{(a^{2} + s^{2})^{2}},
    \label{catemary-potential}
\end{equation}
which is the GIP for a  particle constrained to the catenary (see Fig. \ref{fig:catenarypot}). Thus, from \eqref{curve-schrodinger} we get the Schr\"{o}dinger equation
\begin{equation}
     -\dfrac{\hbar^{2}}{2m}\dfrac{d^{2}\psi}{ds^{2}} -\dfrac{\hbar^{2}}{8m}\dfrac{a^{2}}{(a^{2} + s^{2})^{2}}\psi = E\psi,
    \label{catenary-schrodinger}
\end{equation}
whose solutions for bound states obey the boundary conditions: $\displaystyle{\lim_{s \rightarrow \pm \infty} \psi = 0}$. Due to the presence of the  GIP, finding the eigenfunctions and eigenvalues of \eqref{catenary-schrodinger} is not a simple task. For now, it is more convenient developing the CPF to the helicoid and catenoid, thus, we will return to the problem of confinement to a catenary in section \ref{radial-part}.

\subsection{Helicoid and catenoid}
Substituting \eqref{gauss-mean-helicoid-catenoid} into \eqref{surface-potential} we can obtain the GIP for both helicoid and catenoid surfaces,
\begin{equation}
    V_{S}(u) = -\dfrac{\hbar^{2}}{2m}\dfrac{a^{2}}{(a^{2} + u^{2})^{2}},
    \label{geopothelicatenoid}
\end{equation}
which is independent of $\phi$ and has the same behavior as the potential in Fig. \ref{fig:catenarypot} (up to the multiplicative factor $1/4$). From Eqs. \eqref{metric-tensor} and \eqref{metric-tensor-inverse}, we find $g^{11} = \dfrac{1}{a^{2} + u^{2}},$ $g^{12} = g^{21} = 0$, $g^{22} = 1$, and $g = a^{2} + u^{2}$, consequently, the Laplace-Beltrami operator is
\begin{equation}
    \Delta_{g} = \dfrac{1}{a^{2} + u^{2}}\dfrac{\partial^{2}}{\partial \phi^{2}} + \dfrac{\partial^{2}}{\partial u^{2}} + \dfrac{u}{a^{2} + u^{2}}\dfrac{\partial}{\partial u},
    \label{lapbelheliccat}
\end{equation}
from \eqref{LBop}.
Using Eqs. \eqref{geopothelicatenoid} and \eqref{lapbelheliccat} in Eq. \eqref{schrodinger-surface}, we get the Schr\"{o}dinger equation
\begin{equation}
\begin{split}
    -\dfrac{\hbar^{2}}{2m} \bigg(\dfrac{1}{a^{2} + u^{2}}\dfrac{\partial^{2}\chi}{\partial \phi^{2}} + \dfrac{\partial^{2}\chi}{\partial u^{2}} + \dfrac{u}{a^{2} + u^{2}}\dfrac{\partial\chi}{\partial u}\bigg) & \\
    -\dfrac{\hbar^{2}}{2m}\dfrac{a^{2}}{(a^{2} + u^{2})^{2}}\chi & = E\chi,
    \label{schrodinger-helicatenoid}
    \end{split}
\end{equation}
for the wave function $\chi = \chi(\phi,u)$.

With the element of area given by $dA = \sqrt{g} d\phi du$, the probability of finding the particle in a given region of the surface is 
\begin{equation}
    \begin{split}
    P_{A} & = \iint |\chi(\phi, u)|^{2} (a^{2} + u^{2})^{1/2} d\phi du \\
        & = \iint |\chi_{g}(\phi, u)|^{2} d\phi du,
    \end{split}
    \label{probarea}
\end{equation}
where  (see \cite{kar1994scalar,silva2020electronic}) $\chi_{g}(\phi,u) = (a^{2} + u^{2})^{1/4} \chi(\phi, u)$. Thus, Eq. \eqref{schrodinger-helicatenoid} becomes
\begin{equation}
\begin{split}
        -\dfrac{\hbar^{2}}{2m}\bigg(\dfrac{1}{a^{2} + u^{2}}\dfrac{\partial^{2}\chi_{g}}{\partial\phi^{2}} + \dfrac{\partial^{2}\chi_{g}}{\partial u^{2}}\bigg) & \\ 
        - \dfrac{\hbar^{2}}{2m}\bigg(\dfrac{a^{2}}{4(a^{2} + u^{2})^{2}} + \dfrac{1}{4(a^{2} + u^{2})}\bigg)\chi_{g} & = E\chi_{g}.
        \label{new-schrodinger-helicatenoid}
    \end{split}
\end{equation}
Considering a separation of variables for the wave function, $\chi_{g}(\phi, u) = \Phi(\phi) \psi(u)$, we  write from \eqref{new-schrodinger-helicatenoid} an angular equation,
\begin{equation}
    \dfrac{d^{2}\Phi}{d\phi^{2}} + l^{2}\Phi = 0,
    \label{schrodinger-phi-helicatenoid}
\end{equation}
and a radial equation,
\begin{equation}
  \begin{split}
        -\dfrac{\hbar^{2}}{2m} \dfrac{d^{2}\psi}{du^{2}} - \dfrac{\hbar^{2}}{2m} \bigg( & \dfrac{a^{2}}{4(a^{2} + u^{2})^{2}}   \\ 
    & + \dfrac{1 - 4l^{2}}{4(a^{2} + u^{2})} \bigg) \psi = E\psi, \label{schrodinger-u-helicatenoid}
\end{split}  
\end{equation}
where $l^{2}$ is a separation constant associated with the angular momentum of the particle. 

As we can see, Eq. \eqref{schrodinger-phi-helicatenoid} admits the normalized solutions
\begin{equation}
    \Phi_{l}(\phi) = \dfrac{1}{\sqrt{2\pi}} e^{il\phi}
    \label{angular-catenoid}
\end{equation}
for the catenoid. Thus, in this specific case, it is clear that the parametrization of the catenoid is periodic in $\phi$, because we have $\mathbf{r}(\phi + 2\pi,u) = \mathbf{r}(\phi,u)$, such that the angular wave function must also have periodicity $2 \pi$, \textit{i.e.},  $\Phi_{l}(\phi + 2\pi) = \Phi_{l}(\phi)$, which implies that the angular momentum quantum number $l$ assumes only integer values ($l = 0, \pm 1, \pm 2, \dots$).

The angular wave functions in the case of the helicoid behave similarly to the solutions found in the case of the helix. Due to the geometry of the infinite helicoid ($ \phi \in \mathbb{R}$), equation \eqref{angular-catenoid} is obtained for $l = 0, \pm 1, \pm 2, \dots$, exactly as in the case of the catenoid. On the other hand, for a finite helicoid at $0 < \phi < \phi_{0}$, we find the wave functions
\begin{equation}
    \Phi_{j}(\phi) = \sqrt{\dfrac{2}{\phi_{0}}} \sin\left(\dfrac{j\pi}{\phi_{0}} \phi \right),
    \label{helicoid-finite}
\end{equation}
for $ l = \dfrac{j \pi}{\phi_{0}}$, where $j = 1, 2, 3, \dots$ is the quantum number. Thus, despite the well-known isometry between a catenoid and a  helicoid limited by $ 0 < \phi < 2 \pi$, the respective wave functions differ due to the boundary conditions employed. Thereby, in section \ref{radial-part} we will look for solutions of the radial equation \eqref{schrodinger-u-helicatenoid} for possible bound states, \textit{i.e.}, considering the usual boundary condition $\displaystyle{\lim_{u \rightarrow \pm \infty} \psi = 0}$.

\section{Radial Wave Functions}
\label{radial-part}
As we can see, Eq. \eqref{schrodinger-u-helicatenoid} can be conveniently rewritten as
\begin{equation}
    -\dfrac{d^{2}\psi}{d\tilde{u}^{2}} -\left(\dfrac{1}{4(1 + \tilde{u}^{2})^{2}} + \dfrac{1 - 4l^{2}}{4(1 + \tilde{u}^{2})}\right)\psi = \epsilon\psi,
    \label{adimentionaldiff}
\end{equation}
where $\tilde{u} = u/a$ is the radial variable and $\epsilon = 2ma^{2}E/\hbar^{2}$ is the energy, both in dimensionless units \cite{dandoloff2010geometry}. Thus, \eqref{adimentionaldiff} is a one-dimensional Schr\"{o}dinger equation with an effective potential (also dimensionless) given by
\begin{equation}
    V_{l}(\tilde{u}) = -\dfrac{1}{4(1 + \tilde{u}^{2})^{2}} - \dfrac{1 - 4l^{2}}{4(1 + \tilde{u}^{2})},
    \label{effpot}
\end{equation}
which is always attractive only if $|l| \leq 1/2$. Effective potentials which are equivalent to \eqref{effpot} were previously found in Refs. \cite{atanasov2009geometry, dandoloff2010geometry, da2017quantum, silva2020electronic}.

We see that Eq. \eqref{schrodinger-u-helicatenoid} becomes exactly \eqref{catenary-schrodinger} when considering $l = 1/2$ and the arc length $s = u$. Since $l$ is associated with the angular momentum, it does not appear in the CPF when considering a particle in the catenary, but $l = 1/2$ can be used as a mere mathematical artifice such that the solutions of the Schrödinger equation for the catenary can be obtained through Eq. \eqref{schrodinger-u-helicatenoid}.

\begin{figure}[htp!]
    \centering
    \includegraphics[width=0.9\columnwidth]{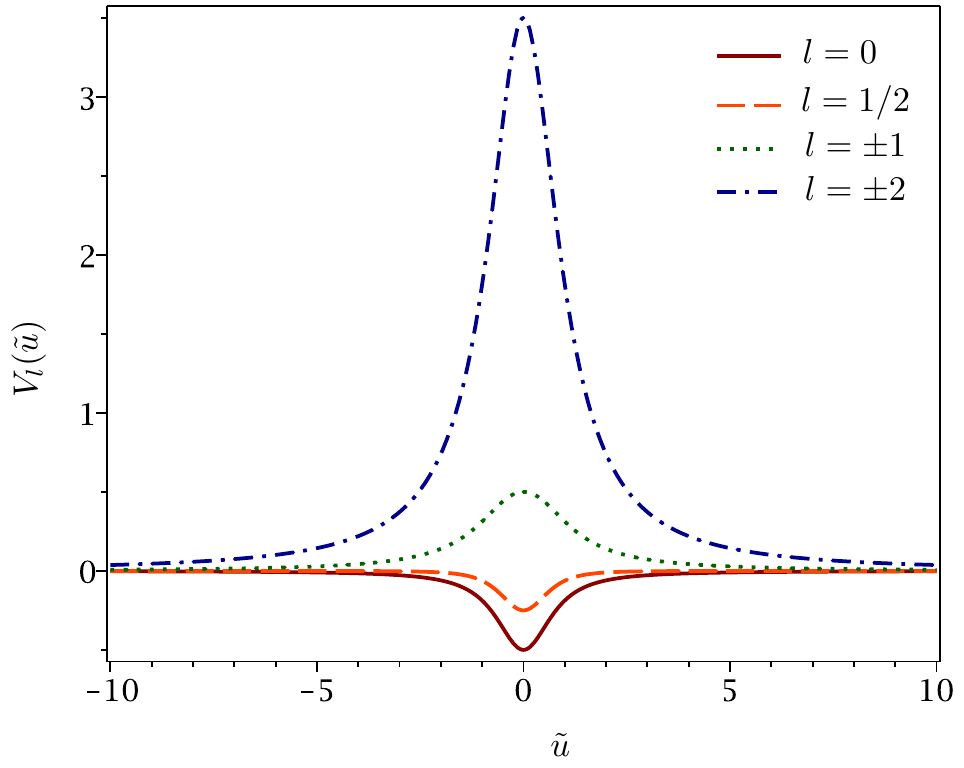}
    \caption{Effective potential $V_{l}(\tilde{u})$ for $l = 0, 1/2, \pm 1, \pm 2$. When $|l|$ increases, $V_{l}(\tilde{u})$ becomes increasingly repulsive.}
    \label{fig:effectiveplot}
\end{figure}

On the other hand, for the finite helicoid we have that $l = \dfrac{j \pi}{\phi_{0}}$, such that the effective potential in \eqref{effpot} is always attractive for $j \leq \dfrac{\phi_{0}}{2\pi}$. Thus, if $\phi_{0}$ is a multiple of $ 2\pi $, the maximum value of $j$ for which $V_{l}(\tilde{u})$ is always attractive is precisely the number of complete loops of the helicoid. Due to the local isometry between the catenoid and the helicoid,  we will discuss only the case where $\phi_{0} = 2\pi$, which implies that $V_{l}(\tilde{u})$ is attractive only for $j = 1$, \textit{i.e.}, for $l = 1/2$. Finally, for the infinite helicoid and the catenoid we have $l = 0, \pm 1, \pm 2, \dots$, thus in these cases the effective potential is always attractive only if $l = 0$, \textit{i.e.}, when the particle has zero angular momentum.  Concisely, to find the possible bound states due to the effective potential we will consider in \eqref{effpot} that
\begin{equation}
    l =
    \begin{cases} 
        1/2  \text{\space for the catenary,} \\
        1/2 \text{\space for the finite helicoid } (\phi_{0} = 2\pi), \\
        0 \text{\space for the catenoid,} \\
        0 \text{\space for the infinite helicoid \space} (\phi \in \mathbb{R}).
   \end{cases}
   \label{lcases}
\end{equation}
In Fig. \ref{fig:effectiveplot} we present the graphs of the effective potential $V_{l}(\tilde{u})$ for some values of $l$.

The second term in the effective potential \eqref{effpot}  can be separated in both repulsive and attractive parts, given by $\dfrac{l^{2}}{1 + \tilde{u}^{2}}$ and  $- \dfrac{1}{4(1 + \tilde{u}^{2})}$, respectively. The repulsive part, which naturally disappears for $l = 0$, corresponds to the usual centrifugal term in quantum mechanics. On the other hand, the attractive one arises due to the confinement of the particle, and it is related to the change made in Eq. \eqref{probarea}, being associated with the so-called ``quantum anticentrifugal force'', as discussed in Refs. \cite{cirone2001quantum,atanasov2007curvature}.

\subsection{Variational method}
Considering the Hamiltonian in Eq. \eqref{adimentionaldiff},
\begin{equation}
    \hat{\mathcal{H}} =  -\dfrac{d^{2}}{d\tilde{u}^{2}} - \left(\dfrac{1}{4(1 + \tilde{u}^{2})^{2}} + \dfrac{1 - 4l^{2}}{4(1 + \tilde{u}^{2})}\right),
\end{equation}
we can use the variational method (also known as the Rayleigh-Ritz method) to obtain approximations for the energy. For a given tentative wave function $\psi_{n,l} = \psi_{n,l}(\tilde{u})$, which is normalized,  the expectation value of $\hat{\mathcal{H}}$ is
\begin{equation}
    \begin{split}
        \epsilon_{n,l} = & - \int_{-\infty}^{+ \infty} \psi_{n,l}^{*}\dfrac{d^{2}\psi_{n,l}}{d\Tilde{u}^{2}}d\tilde{u} \\
        & - \dfrac{1}{4}\int_{-\infty}^{+ \infty}\dfrac{\psi_{n,l}^{*}\psi_{n,l}}{(1 + \tilde{u}^{2})^{2}} d\tilde{u} \\
        & - \dfrac{1 - 4l^{2}}{4}\int_{-\infty}^{+ \infty} \dfrac{\psi_{n,l}^{*}\psi_{n,l}}{1 + \tilde{u}^{2}} d\tilde{u},
    \label{H-expectation}
    \end{split}
\end{equation}
where $n = 1, 2, 3, \dots$ is the quantum number associated with the energy levels. Such method consists of intuitively proposing a wave function $\psi_{n,l}$ which depends on a parameter $\beta_{l}$ adjusted in order to minimize the energy $\epsilon_{n,l}(\beta_{l})$ in Eq. \eqref{H-expectation}. Consequently, this method give us an upper limit for the true energy \cite{zettili2009concepts,sakurai_napolitano_2017}. 

\subsubsection{Gaussian function}
For the ground state ($n = 1$) we are looking for bell-shaped functions. We firstly propose the Gaussian wave function
\begin{equation}
    \psi_{1,l}(\tilde{u}) = \left(\dfrac{\beta_{l}}{\pi}\right)^{1/4}e^{-\beta_{l} \tilde{u}^{2}/2},
    \label{wavetest}
\end{equation}
where $\beta_{l} > 0 $ is a parameter to be adjusted in order to obtain a minimum value for the energy of the particle. By replacing \eqref{wavetest} in \eqref{H-expectation}, we obtain
\begin{equation}
    \begin{aligned}
        \epsilon_{1,l} = - \sqrt{\dfrac{\beta_{l}}{\pi}} \Bigg[ & \beta_{l} \int_{-\infty}^{+ \infty} \left(\beta_{l} \tilde{u}^{2} - 1\right)e^{-\beta_{l} \tilde{u}^{2}} d\tilde{u}  \\
         & + \dfrac{1}{4}\int_{-\infty}^{+ \infty}\dfrac{e^{-\beta_{l} \tilde{u}^{2}}}{(1 + \tilde{u}^{2})^{2}} d\tilde{u} \\
         & +  \dfrac{1-4l^{2}}{4}\int_{- \infty}^{+ \infty} \dfrac{e^{-\beta_{l} \tilde{u}^{2}}}{1 + \tilde{u}^{2}} d\tilde{u} \Bigg],
        \label{gauss-energy}
    \end{aligned}
\end{equation}
which leads us to
\begin{equation}
    \begin{split}
       \epsilon_{1,l} = \ & \dfrac{\sqrt{\pi\beta_{l}}e^{\beta_{l}}}{4}\left(\beta_{l} + 4l^{2} - \dfrac{3}{2}\right)\left[1 - \erf\left(\sqrt{\beta_{l}}\right)\right] \\
        & + \dfrac{\beta_{l}}{4},
\end{split}
   \label{energybeta-gauss}
\end{equation}
where $\erf$ is the error function \cite{jeffrey2007table,olver2010nist}. By minimizing the energy with respect to $\beta_{l}$, we find
\begin{equation}
    \begin{split}
     \dfrac{\sqrt{\pi}e^{\beta_{l}}}{4\sqrt{\beta_{l}}} \left[\beta_{l}^{2} + 2l^{2}\left(2\beta_{l} + 1\right)-\dfrac{3}{4}\right]\left[1 - \erf\left(\sqrt{\beta_{l}}\right)\right] & \\
        + \dfrac{5}{8} -l^{2} -\dfrac{\beta_{l}}{4} = \ & 0.
    \end{split}
    \label{diffenergy-gauss}
\end{equation}

\begin{figure}[htp!]
     \centering
     \includegraphics[width=1\columnwidth]{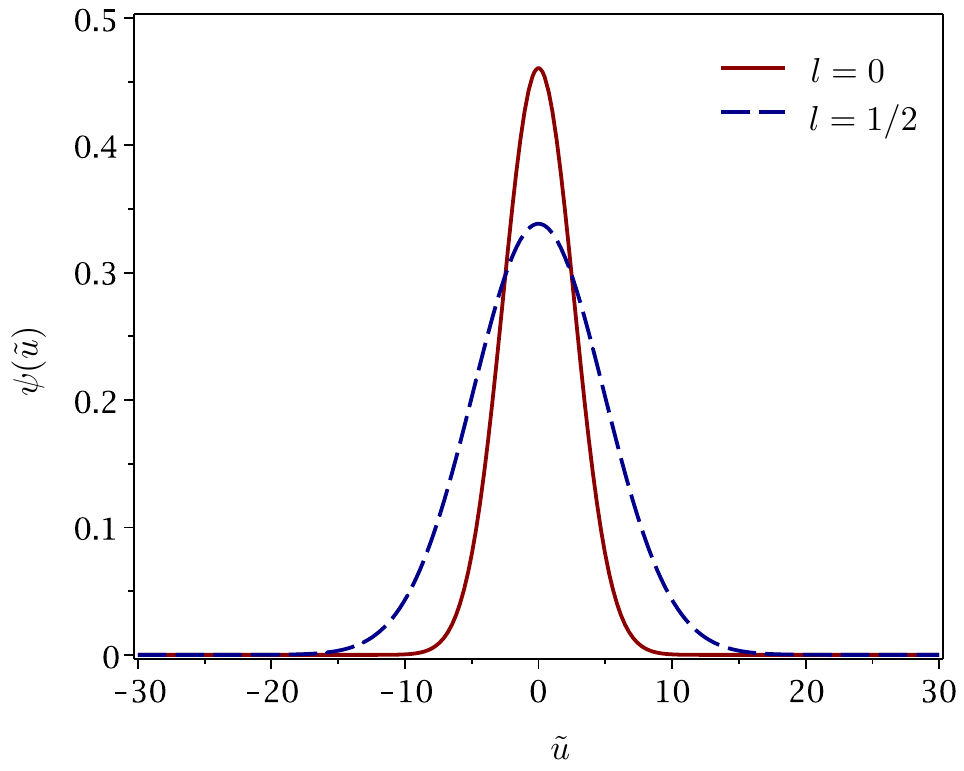}
     \caption{Gaussian wave functions obtained with the variational method for the ground state considering $l = 0$ for the infinite helicoid and the catenoid, as well as $l = 1/2$ in for the finite helicoid ($\phi_{0} = 2\pi$) and the catenary.}
     \label{fig:gauss-var-meth}
 \end{figure}

Now, we need to solve Eq. \eqref{diffenergy-gauss} numerically for the corresponding values of $l$. For the infinite helicoid and for the catenoid we have $l = 0$, such that \eqref{diffenergy-gauss} becomes
\begin{equation}
    \dfrac{\sqrt{\pi}e^{\beta_{0}}}{4\sqrt{\beta_{0}}} \left(\beta_{0}^{2} -\dfrac{3}{4}\right) \left[1 - \erf\left(\sqrt{\beta_{0}}\right)\right] + \dfrac{5}{8} -\dfrac{\beta_{0}}{4} = 0,
\end{equation}
 whose approximate solution is $\beta_{0} \approx 0.14133$, therefore, we get the energy $\epsilon_{1,0} \approx - 0.11977$. On the other hand, we have $l = 1/2$ in the cases of the finite helicoid ($\phi_{0} = 2\pi$) and the catenary, such that we find
\begin{equation}
    \begin{split}
        \dfrac{\sqrt{\pi}e^{\beta_{1/2}}}{4\sqrt{\beta_{1/2}}} \left(\beta_{1/2}^{2} + \beta_{1/2} - \dfrac{1}{4}\right) \left[1 - \erf\left(\sqrt{\beta_{1/2}} \ \right)\right] & \\
    + \dfrac{3}{8} -\dfrac{\beta_{1/2}}{4} = \ & 0,
    \end{split}
\end{equation}
thus $\beta_{1/2} \approx 0.04120$ and $\epsilon_{1,1/2} \approx -0.02299$. Fig. \ref{fig:gauss-var-meth} shows the graphs of the corresponding tentative wave functions from Eq. \eqref{wavetest}.
 
\subsubsection{Lorentzian function}
Also for the ground state, we propose (for comparison purposes) the Lorentzian wave function
 \begin{equation}
    \psi_{1,l}(\tilde{u}) = \left(\dfrac{2\beta_{l}^{3}}{\pi}\right)^{1/2} \dfrac{1}{\beta_{l}^{2} + \tilde{u}^{2}},
    \label{wavetest2}
\end{equation}
 where $\beta_{l} > 0$. By replacing Eq. \eqref{wavetest2} in \eqref{H-expectation}, we obtain
\begin{equation}
    \begin{split}
        \epsilon_{1.l} = -\dfrac{2\beta_{l}^{3}}{\pi} & \Bigg[ 2\int_{-\infty}^{+ \infty} \dfrac{ 3\tilde{u}^{2} - \beta_{l}^{2}}{(\beta_{l}^{2} + \tilde{u}^{2})^{4}} d\tilde{u} \\
        & + \dfrac{1}{4} \int_{-\infty}^{+ \infty}\dfrac{d\tilde{u}}{(1 + \tilde{u}^{2})^{2}(\beta_{l}^{2} + \tilde{u}^{2})^{2}} \\
        & +\dfrac{1 - 4l^{2}}{4}\int_{-\infty}^{+ \infty} \dfrac{d\tilde{u}}{(1 + \tilde{u}^{2})(\beta_{l}^{2} + \tilde{u}^{2})^{2}} \Bigg].
    \end{split}
    \label{epsilon-lorentz}
\end{equation}
From equation \eqref{epsilon-lorentz} we get
\begin{equation}
    \epsilon_{1,l} = \dfrac{1}{2\beta_{l}^{2}} - \dfrac{\beta_{l}^{2} + 3\beta_{l} + 1}{4(\beta_{l} + 1)^{3}} + \dfrac{4l^{2} - 1}{4}\dfrac{2\beta_{l} + 1}{(\beta_{l} + 1)^{2}},
\end{equation}
whose minimization gives us
\begin{equation}
    \begin{split}
    (3 - 8l^{2})\beta_{l}^{5} + \left(2 - 8l^{2}\right)\beta_{l}^{4} -16\beta_{l}^{3} & \\
    - 24\beta_{l}^{2} -16\beta_{l} - 4 & = 0.
    \label{diffenergy-lorentz}
    \end{split}
\end{equation}
 
\begin{figure}[htp!]
     \centering
     \includegraphics[width=1\columnwidth]{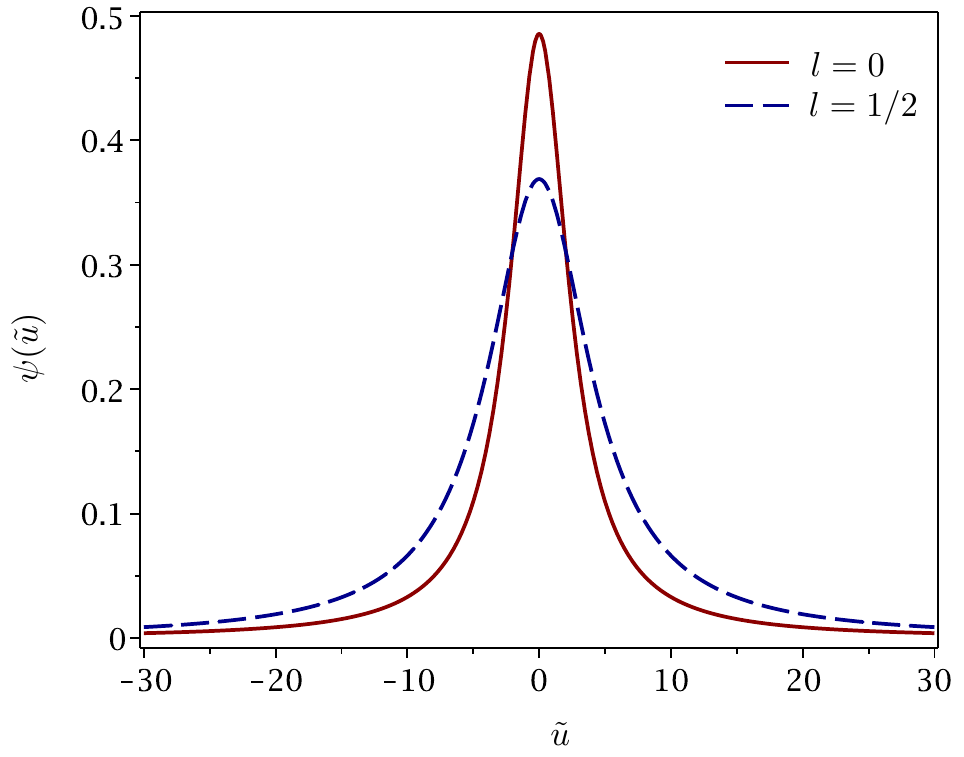}
     \caption{Lorentzian wave functions obtained with the variational method for the ground state considering $l = 0$ for the infinite helicoid and the catenoid, as well as $l = 1/2$ for the finite helicoid ($\phi_{0} = 2\pi$) and the catenary.}
     \label{fig:lorentz-var-meth}
 \end{figure}

 Thus, for the infinite helicoid and for the catenoid we set $l = 0$ in Eq. \eqref{diffenergy-lorentz}, which give us
\begin{equation}
    3\beta_{0}^{5} + 2\beta_{0}^{4} -16\beta_{0}^{3} - 24\beta_{0}^{2} - 16\beta_{0} - 4 = 0,
\end{equation} 
whose positive solution is approximately $\beta_{0} \approx 2.70072$, and from that, we get the energy $\epsilon_{1,0} \approx -0.12918$. Similarly, we have $l = 1/2$ in the cases of the finite helicoid ($\phi_{0} = 2\pi$) and of the catenary, thus
\begin{equation}
    \beta_{1/2}^{5} -16\beta_{1/2}^{3} - 24\beta_{1/2}^{2} - 16\beta_{1/2} - 4 = 0,
\end{equation}
therefore $\beta_{1/2} \approx 4.67957$ and $\epsilon_{1,1/2} \approx -0.02757$. Figure \ref{fig:lorentz-var-meth} shows the graphs of the tentative wave functions obtained with the respective $\beta_{l}$.

 As we can see with the Gaussian and Lorentzian functions, the wave function is more concentrated at the origin for $l = 0$ than for $l = 1/2$. This was expected, as the second term in Eq. \eqref{effpot} cancels out for $l = 1/2$, but is negative for $l = 0$, making the particle in the infinite helicoid or in the catenoid feel a more attractive effective potential (see Fig. \ref{fig:effectiveplot}). Furthermore, in both cases, the respective energies found with the Gaussian function are greater than those found with the Lorentzian function, which indicates that Eq. \eqref{wavetest2} is a better approximation for the wave function, since the variational method always provides an upper limit for the exact energy of the particle \cite{sakurai_napolitano_2017}.
 
\subsection{Confluent Heun function}
The variational method provides reasonable approximations for the ground state of the problems in question, but it does not necessarily works well in the search for possible excited states. Consequently, we need to turn our attention again to the radial equation, 
\begin{equation}
    \dfrac{d^{2}\psi}{d\tilde{u}^{2}} + \left(\dfrac{1}{4(1 + \tilde{u}^{2})^{2}} + \dfrac{1 - 4l^{2}}{4(1 + \tilde{u}^{2})} + \epsilon\right)\psi = 0,
    \label{radialeq2}
\end{equation}
in order to find exact solutions or better approximations.

Introducing a new variable  $\xi = - \tilde{u}^{2}$, such that $\xi \in (-\infty, 0],$ we find from Eq. \eqref{radialeq2} that
\begin{equation}
    \begin{split}
         & \dfrac{d^{2}\psi}{d\xi^{2}} + \dfrac{1}{2\xi}\dfrac{d\psi}{d\xi} \\
         & -\dfrac{4\epsilon\xi^{2} + (4l^{2} -8\epsilon -1)\xi + 4\epsilon -4l^{2} +2 }{16\xi(1 - \xi)^{2}} \psi = 0.
     \label{radialeq3}
\end{split}
\end{equation}
Considering similar changes to those used in Refs. \cite{hartmann2014bound,hartmann2017two,ishkhanyan2016schrodinger}, we use the \textit{ansatz} $\psi(\xi) = (1 - \xi)^{(\tilde{\gamma} + 1)/2}H(\xi)$, where $\tilde{\gamma} > 0$ is a parameter to be conveniently adjusted.
From \eqref{radialeq3} we get
\begin{equation}
    \dfrac{d^{2}H}{d\xi^{2}} + \left[ \dfrac{1}{2\xi} - \dfrac{\tilde{\gamma} + 1}{1 - \xi} \right] \dfrac{dH}{d\xi} + \dfrac{ \mathcal{A}(\xi) }{16\xi(1 - \xi)^{2}}H = 0,
    \label{radialeq4}
\end{equation}
where we have the function
\begin{equation*}
    \mathcal{A}(\xi) = (1 - \xi)\left[4l^{2} - 4\tilde{\gamma}(\tilde{\gamma} +1) -1 -4\epsilon(1 - \xi) \right] + 4{\tilde{\gamma}}^{2} - 5. 
\end{equation*}

Since $\tilde{\gamma} $ is ``arbitrary'', we can consider it in a way that Eq. \eqref{radialeq4} becomes as simple as possible. Thus, choosing  $\tilde{\gamma}$ such that $4{\tilde{\gamma}}^{2} - 5 = 0$ is the most convenient here, because these are precisely the terms that are not being multiplied by $(1 - \xi)$ in $\mathcal{A}(\xi)$. Therefore, we have the parameter $\tilde{\gamma} = \sqrt{5}/2$, thus it follows from Eq. \eqref{radialeq4} that
\begin{equation}
    \begin{split}
        \dfrac{d^{2}H}{d\xi^{2}} + \left[ \tilde{\alpha} + \dfrac{\tilde{\beta} + 1}{\xi} + \dfrac{\tilde{\gamma} + 1}{\xi - 1} \right] \dfrac{dH}{d\xi} & \\
         + \left[ \dfrac{\tilde{\mu}}{\xi} + \dfrac{\tilde{\nu}}{\xi - 1} \right]H & = 0,
    \label{heun-confluent-eq}
    \end{split}
\end{equation}
for $\tilde{\alpha} = 0$, $\tilde{\beta} = - 1/2$, $\tilde{\mu} = \left(2l^{2} - \sqrt{5} -3 -2\epsilon \right)/8$, and $\tilde{\nu} = \left(\sqrt{5} +3 -2l^{2}\right)/8$. Equation \eqref{heun-confluent-eq} is known as the confluent Heun equation \cite{karayer2015extension,arscott1995heun,ishkhanyan2016schrodinger,kristensson2010second,ishkhanyan2018schrodinger}. 

Around the regular singular point $\xi = 0$, the power series solution of Eq. \eqref{heun-confluent-eq} converges on the unit disc, $|\xi| < 1$, and it is called the confluent Heun function,
\begin{equation}
    \HeunC\left(\tilde{\alpha}, \tilde{\beta}, \tilde{\gamma}, \tilde{\delta}, \tilde{\eta}, \xi\right) = \sum_{n = 0}^{\infty} v_{n}\left(\tilde{\alpha}, \tilde{\beta}, \tilde{\gamma}, \tilde{\delta}, \tilde{\eta}\right)\xi^{n},
    \label{heunc-func}
\end{equation}
with the parameters
\begin{align}
    \tilde{\delta}  & =  \tilde{\mu} + \tilde{\nu} -\dfrac{\tilde{\alpha}}{2}\left(\tilde{\beta} + \tilde{\gamma} + 2\right) = -\dfrac{\epsilon}{4},\\
    \tilde{\eta} & = \dfrac{\tilde{\alpha}}{2}\left(\tilde{\beta} + 1\right) - \tilde{\mu} -\dfrac{\tilde{\beta} + \tilde{\gamma} +\tilde{\beta}\tilde{\gamma}}{2} = \dfrac{2\epsilon - 2l^{2} + 5}{8},
\end{align}
and where $v_{n} = v_{n}\left(\tilde{\alpha}, \tilde{\beta}, \tilde{\gamma}, \tilde{\delta}, \tilde{\eta}\right)$ are coefficients \cite{karayer2015extension,downing2013solution}. Even though the convergence of \eqref{heunc-func} is assured for $|\xi| < 1$, in many cases it is possible to use numerical methods of extrapolation (like those implemented in Maple) to obtain approximations of $H(\xi)$ when $|\xi| \geq 1$, depending on the parameters $\tilde{\alpha}$, $\tilde{\beta}$, $\tilde{\gamma}$, $ \tilde{\delta}$, and $\tilde{\eta}$.

Since $\tilde{\beta} $ is non-integer, a second solution for \eqref{heun-confluent-eq} can be written as $\xi^{-\tilde{\beta}} \HeunC\left(\tilde{\alpha}, -\tilde{\beta}, \tilde{\gamma}, \tilde{\delta}, \tilde{\eta}, \xi\right)$ \cite{boyack2011confluent,olver2010nist}. Thus, from \eqref{heunc-func}, the solutions of Eq. \eqref{radialeq2} are separable into an even function,
\begin{equation}
    \psi_{\text{e}}(\tilde{u}) = C_{0}\left(1 + \tilde{u}^{2}\right)^{\frac{\sqrt{5}}{4} + \frac{1}{2}}  \HeunC\left(\tilde{\alpha}, \tilde{\beta}, \tilde{\gamma}, \tilde{\delta}, \tilde{\eta}, -\tilde{u}^{2}\right),
    \label{sol-even}
\end{equation}
and an odd function,
\begin{equation}
    \psi_{\text{o}}(\tilde{u}) = C_{1}\left(1 + \tilde{u}^{2}\right)^{\frac{\sqrt{5}}{4} + \frac{1}{2}} \tilde{u} \HeunC\left(\tilde{\alpha}, - \tilde{\beta}, \tilde{\gamma}, \tilde{\delta}, \tilde{\eta}, -\tilde{u}^{2}\right),
     \label{sol-odd}
\end{equation}
where $C_{0}$ and $C_{1}$ are normalization constants.

Once the analytical solutions of \eqref{radialeq2} have been found, the next step is to apply the boundary conditions $\displaystyle{\lim_{\tilde{u} \rightarrow \pm \infty} \psi = 0}$. We cannot use here the approach of reducing $\HeunC$ to a polynomial (see \cite{arscott1995heun,downing2013solution,fiziev2009novel}), because we would get solutions that do not obey the boundary conditions. Similar problems have been reported in some recent works \cite{dong2016exact,dong2016semi,dong2018semiexact,dong2019exact}. An alternative would be writing a solution to  \eqref{heun-confluent-eq} around the regular singular point $\xi = 1$ and use the so-called Wronskian method to obtain the corresponding eigenvalues \cite{hartmann2014bound, boyack2011confluent,fernandez2011wronskian}. However, here we have the variable $\xi = - \tilde{u}^{2}$, which implies that $\xi = 1$ is not part of the solutions domain of Eq. \eqref{heun-confluent-eq} in the problem of interest, therefore, the Wronskian method cannot be applied here.

To circumvent the above mentioned adversities,  we propose that the particle is initially in a system limited in $\tilde{u}$, \textit{i.e.}, subject to the potential 
\begin{equation}
    V(\tilde{u}) =
    \begin{cases} 
        + \infty  \text{\space for \space} \tilde{u} \leq - \tilde{u}_{0} \text{\space or\space} \tilde{u} \geq \tilde{u}_{0}, \\
        V_{l}(\tilde{u}) \text{\space for } - \tilde{u}_{0} < \tilde{u} < \tilde{u}_{0},
   \end{cases}
\end{equation}
where $V_{l}(\tilde{u})$ is the effective potential in \eqref {effpot} and $\pm\tilde{u}_{0}$ corresponds to the edges of the one-dimensional box, whose width is $2\tilde{u}_{0}$. Thus, when we have finite catenaries, helicoids, or catenoids in $\tilde{u}$, both equations 
\begin{equation}
    \lim_{\tilde{u} \rightarrow \pm \tilde{u}_{0}} \HeunC\left(\tilde{\alpha}, \tilde{\beta}, \tilde{\gamma}, \tilde{\delta}, \tilde{\eta}, -\tilde{u}^{2}\right) = 0 
    \label{heunlimit1}
\end{equation}
and
\begin{equation}
    \lim_{\tilde{u} \rightarrow \pm \tilde{u}_{0}} \HeunC\left(\tilde{\alpha}, - \tilde{\beta}, \tilde{\gamma}, \tilde{\delta}, \tilde{\eta}, -\tilde{u}^{2}\right) = 0
    \label{heunlimit2}
\end{equation}
  can be used to obtain the energies $\epsilon_{n,l}$ (for $n = 1, 2, 3,\dots$) that cause wave functions to tend to zero at $\pm\tilde{u}_{0}$ \cite{sitnitsky2017exactly}. When assigning values to $\tilde{u}_{0}$, solutions to \eqref{heunlimit1} and \eqref{heunlimit2} can be found using Maple's \textit{fsolve} command and its numeric methods for $|\xi| \geq 1$ \cite{maple}. Next, we will conduct a study of how the energies and the wave functions behave as the box size increases. We  call this  ``the  box method".

\subsubsection{Attractive effective potential}
\begin{figure}[tp]
    \centering
    \includegraphics[width=0.95\columnwidth]{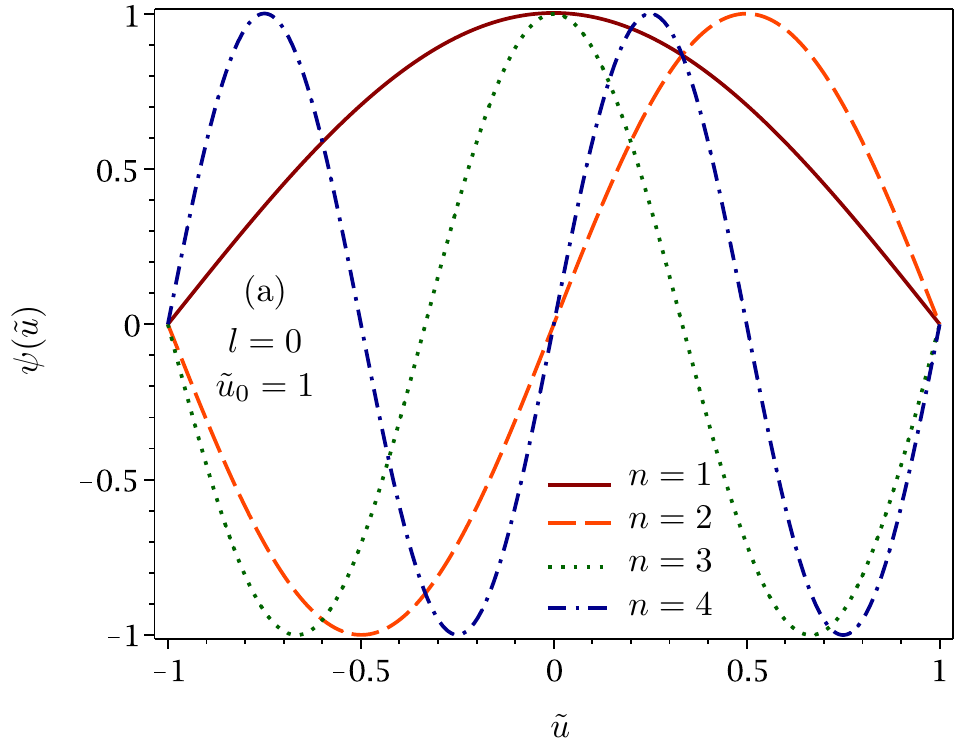} \\
    \vspace{0.525cm}
    \includegraphics[width=0.95\columnwidth]{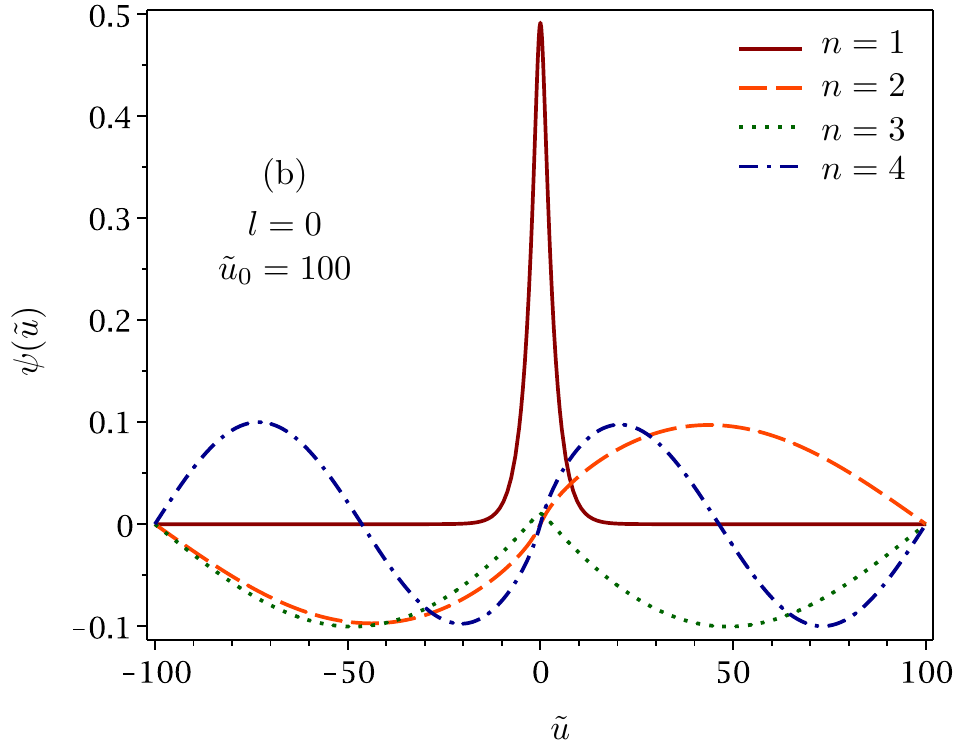} \\
    \vspace{0.525cm}
    \includegraphics[width=0.95\columnwidth]{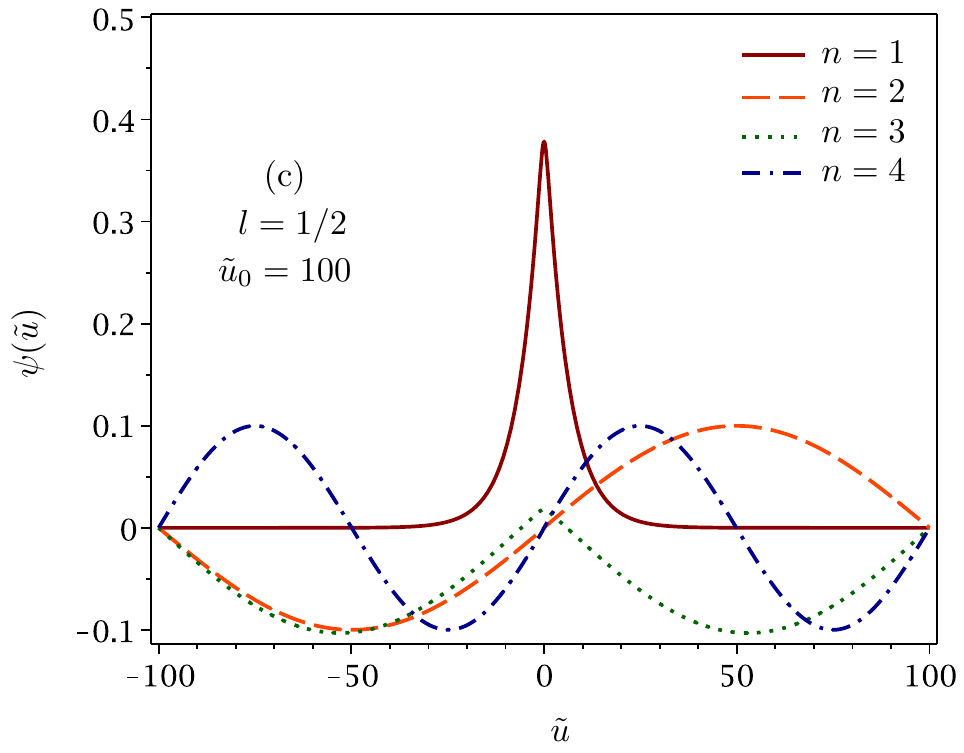}
    \caption{In graph (a) are the radial wave functions (numerically normalized) corresponding to the first four states when considering $\tilde{u}_{0} = 1$ and $l = 0$ (if $\tilde{u}_{0} = 1$ and $l = 1/2$, the solutions behave similarly). In graph (b) are the radial wave functions (numerically normalized) for the first four states when $l = 0$ for $\tilde{u}_{0} = 100$ and, in graph (c), the same when $l = 1/2$.}
    \label{fig:box-1}
\end{figure}

Using the box method, let us discuss here the cases in which the effective potential in Eq. \eqref{effpot} is attractive, \textit{i.e.}, when we set $l = 0$ in the cases of the infinite helicoid and of the catenoid, and $l = 1/2$ in the case of the finite helicoid ($\phi_{0} = 2\pi$) and the catenary, according to Eq. \eqref{lcases}. Table \ref{tab:catenoid-box} shows the energies for some values of $\tilde{u}_{0}$. Thus, we can see that when $\tilde{u}_{0}$ has values around $1$, the effects of the effective potential are not so relevant. For example, when considering $\tilde{u}_{0} = 1$, the energies of the first four states ($n = 1, 2, 3, 4$) when we set $l = 0$ ($ l = 1 / $ 2) are given  approximately by $ 2.03806 $ ($ 2.26282 $), $9.50983$ ($9.70867$), $21.84675$ ($22.04447$) and $39.11990$ ($39.31700$). In this case, the radial wave functions $\psi(\tilde{u})$ for $l = 1/2$ behave similarly to the corresponding wave functions for $l = 0$ (see graph (a) in Fig. \ref{fig:box-1}).

\begin{table}[htp]\centering
    \caption{Energies $\epsilon_{n,l}$ of the first four states ($n = 1, 2, 3, 4$) for $l = 0$ and for $l = 1/2$, considering  increasing values for the box size, $2\tilde{u}_{0}$.}
    \begin{tabular}{ccccc}
        \hline \hline
        $\tilde{u}_{0}$     & $\epsilon_{1,0}$    & $\epsilon_{2,0}$     & $\epsilon_{3,0}$     & $\epsilon_{4,0}$    \\  \hline
        $1$             	& $2.03806$	          & $9.50983$	         & $21.84675$	        & $39.11990$          \\
        $3.00420$           & $0.00000$           & $0.96307$            & $2.27982$            & $4.21282$           \\
        $10$                & $-0.12999$          & $0.07853$            & $0.15781$            & $0.36069$           \\
        $50$                & $-0.13051$          & $0.00289$            & $0.00388$            & $0.01370$           \\
        $100$               & $-0.13051$          & $0.00071$            & $0.00087$            & $0.00338$           \\
        $500$               & $-0.13051$          & $0.00003$            & $0.00003$            & $0.00013$           \\
        $1 000$             & $-0.13051$          & $0.00001$            & $0.00001$            & $0.00001$           \\ \hline \hline
                            &                     &                      &                      &                   \\   \hline \hline 
        $\tilde{u}_{0}$     & $\epsilon_{1,1/2}$   & $\epsilon_{2,1/2}$    & $\epsilon_{3,1/2}$    & $\epsilon_{4,1/2}$     \\  \hline
        $1$                 & $2.26282$	            & $9.70867$	             & $22.04447$	          & $39.31700$              \\ 
        $5.97477$           & $0.00000$             & $0.26695$              & $0.57495$              & $1.08526$               \\ 
        $10$                & $-0.02401$            & $0.09601$              & $0.19112$              & $0.38755$               \\ 
        $50$                & $-0.02892$            & $0.00392$              & $0.00502$              & $0.01568$               \\ 
        $100$               & $-0.02892$            & $0.00098$              & $0.00111$              & $0.00393$               \\ 
        $500$               & $-0.02892$            & $0.00004$              & $0.00004$              & $0.00016$               \\ 
        $1 000$             & $-0.02892$            & $0.00001$              & $0.00001$              & $0.00004$               \\  \hline \hline
        \label{tab:catenoid-box}
    \end{tabular}
    \vspace{0.075cm}
\end{table}

As the box size increases, the energy associated with each level decreases. Thus, the ground state energy of the particle, $\epsilon_{1,l} $, becomes zero for $l = 0$ ($l = 1/2$) when $\tilde{u}_{0} \approx 3.00420$ ($\tilde{u}_{0} \approx 5.97477$), such that if the respective $\tilde{u}_{0}$ is greater than that, $\epsilon_{1,l}$ becomes negative. Therefore, if we have $\tilde{u}_{0} = 100$, the approximate energies of the first four states for $l = 0$ ($l = 1/2$) are $-0.13051$ ($-0.02892 $), $0.00071$ ($0.00098$), $0.00087$ ($0.00111$), and $0.00338$ ($0.00393$), thus, more significant differences between the respective wave functions for $l = 0$ and $l = 1/2$ appear, as shown in graphs (b) and (c) in Fig. \ref{fig:box-1}. Furthermore, in these graphs it is clear that the square module of the respective wave functions is practically the same for $n = 2$ and $n = 3$; this happens for $n = 4$ and $n = 5$ for larger values of $\tilde{u}_{0}$, and so on.

Thus, when the box size increases, the effects of the effective potential on the wave functions becomes increasingly apparent, while the effects due to the box starts to become less relevant. For $l = 0$ ($l = 1/2$), the energy of the ground state $\epsilon_{1.0}$ ($\epsilon_{1,1/2}$) converges to approximately $- 0.13051$ ($ - 0.02892$) as the box size is increased. As we can see in both cases, the respective energies of the following three states decreases when $\tilde{u}_{0}$ increases, in such a way that they become practically indistinguishable from each other. These results show that, for both values of $l$, there will be a state with negative energy starting from a certain value of $\tilde{u}_{0}$, therefore, this is only due to $V_{l}(\tilde{u})$.

\begin{figure}
    \centering
    \includegraphics[width=1\columnwidth]{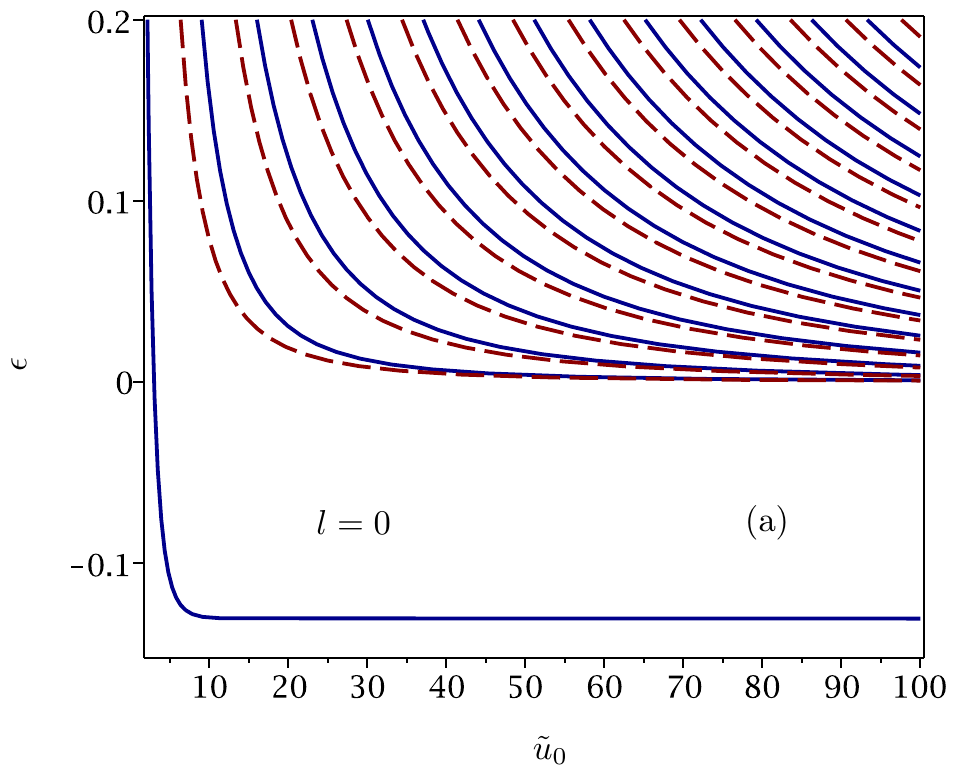} \\
         \vspace{0.3cm}
    \includegraphics[width=1\columnwidth]{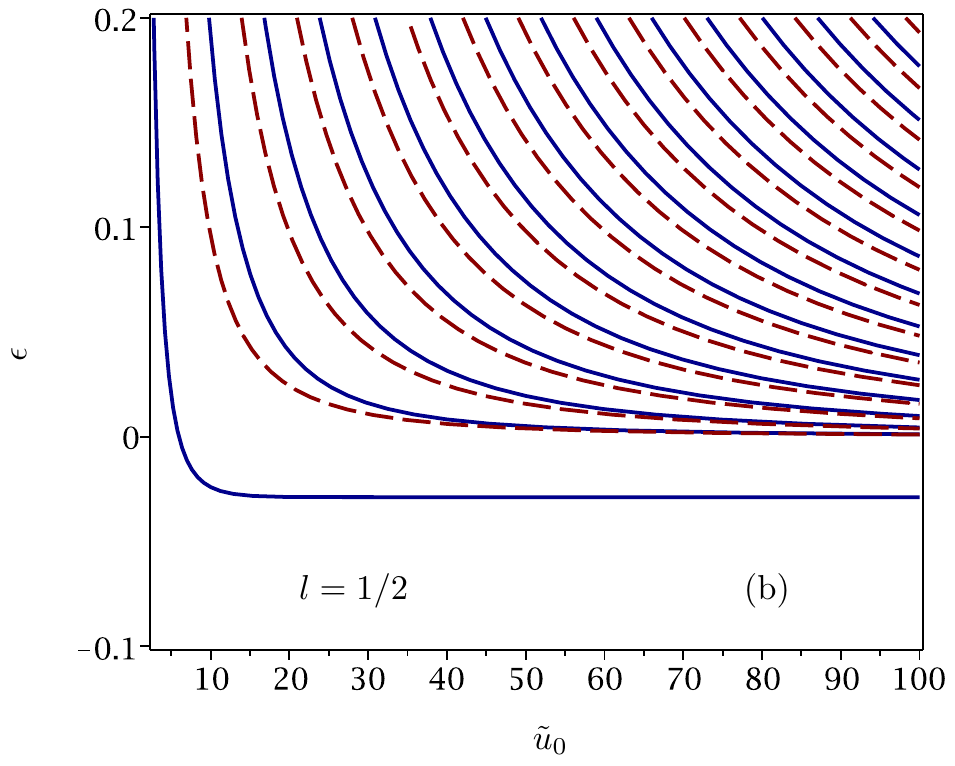}
    \caption{Graph (a) shows the energies as a function of the box size, $2\tilde{u}_{0}$, for $l = 0$ and, in graph (b), the same for $l = 1/2$. Each line corresponds to an energy level, with the blue continuous lines associated with the even solutions, and the red dashed lines with the odd ones.}
    \label{fig:energy-box}
\end{figure}

To better illustrate this, Fig. \ref{fig:energy-box} shows how the energy of the particle varies according to the box size, $2\tilde{u}_{0}$, where the blue continuous (red dashed) lines represent the states with even (odd) wave functions. Thus, the energy of the ground state with $l = 0$ ($l = 1/2$) tends to approximately $- 0.13051$ ($- 0.02892$) as the box size increases, while the energies of the excited states tend to form a continuous spectrum. This leads to the inference that there is only a single bound state for the effective potential presented in equation \eqref{effpot} for the catenary, helicoid, or catenoid with infinite size in $\tilde{u}$.

\begin{figure}[htp!]
        \centering     
        \includegraphics[width=1\columnwidth]{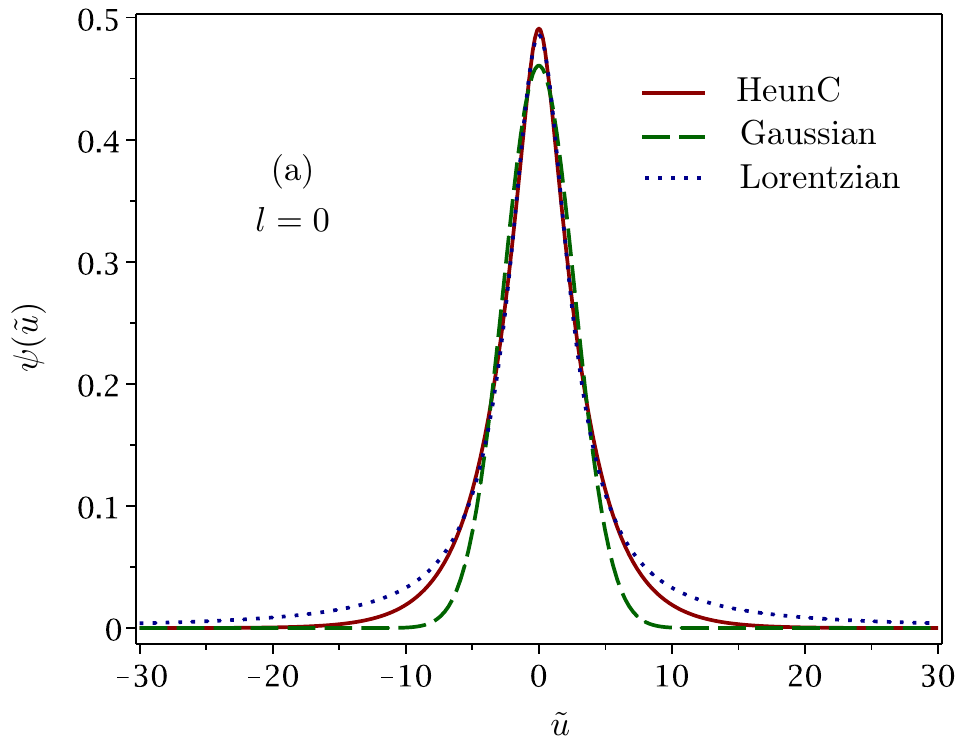} \\
        \vspace{0.3cm}
        \includegraphics[width=1\columnwidth]{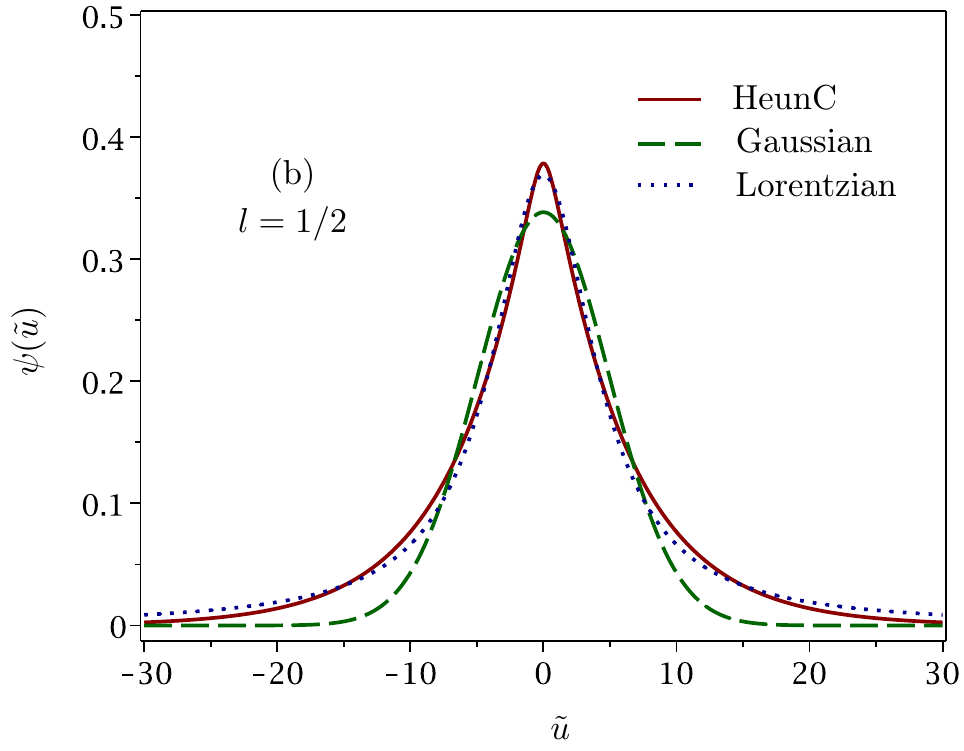}
        \caption{Comparison  between the radial wave functions obtained by the box  method using $\HeunC$, and for the Gaussian and Lorentzian functions used in the variational method (a) for $l = 0$ and (b) for $l = 1/2$.}
        \label{fig:comparasion}
\end{figure}

The ground state energy $\epsilon_{1,l}$ was obtained in an approximate way with the box method for $\tilde{u}_{0} \gg 1$, but the wave functions in \eqref{sol-even} and \eqref{sol-odd} are the exact eigenfunctions of Eq. \eqref{adimentionaldiff}, thus the solution obtained in this work is actually ``semi-exact'' \cite{dong2018semiexact}. Fig. \ref{fig:comparasion} compares these solutions with the wave functions found using the variational method for both $l = 0$ and $l = 1/2$. As one can see, the confluent Heun function goes to zero faster than the Lorentzian, but slower than the Gaussian. For both values of $l$, the respective energy found with the box  method is considerably near those obtained with the variational method, especially for the Lorentzian function.

\subsubsection{Repulsive effective potential}
As we can see, the situations in which the effective potential in \eqref{effpot} is always repulsive correspond to $|l| > 1/2$. Thus, for a helicoid with $\phi_{0} = 2\pi$ this occurs if the quantum number $j > 1 $, \textit{i.e.}, when $l = 1, 3/2, 2, 5/2, \dots$, while it happens for $l = \pm 1, \pm 2, \pm 3, \dots$ in the case of the infinite helicoid and of the catenoid. In these cases all bound states have positive energy and are due only to the box, such that at the limit where $\tilde{u}_{0} \rightarrow \infty$ there will be no bound state for both surfaces. As examples, the cases where $l = \pm 1$ and $l = \pm 2$ for $\tilde{u}_{0} = 1$ and $\tilde{u_{0}}= 100$ are discussed below. For the semi-integer values $l = 3/2, 5/2, \dots$, the results are completely analogous.

When we have $\tilde{u}_{0} = 1$, the energies for the first four states where $l = \pm 1$ ($l = \pm 2$) are given approximately by $2.93669$ ($5.62598$), $10.30488$ ($12.68535$), $22.63779$ ($25.01384$), and $39.90841$ ($42.27562$), whose respective wave functions are also very similar to those shown in graph (a) in Fig. \ref{fig:box-1}. In such cases, this also suggests that, for values of $\tilde{u}_{0}$ near $1$, the effects of the box have more influence on the eigenfunctions than the presence of the effective potential. However, as the values of $|l|$ increase, $V_{l}(\tilde{u})$ becomes more and more repulsive, tending to ``push'' the probability density $|\psi|^{2}$ further away from the origin, even for relatively small $\tilde{u}_{0} $ values, \textit{i.e.}, near $1$.

\begin{figure}[h]
        \centering 
        \includegraphics[width=0.95\columnwidth]{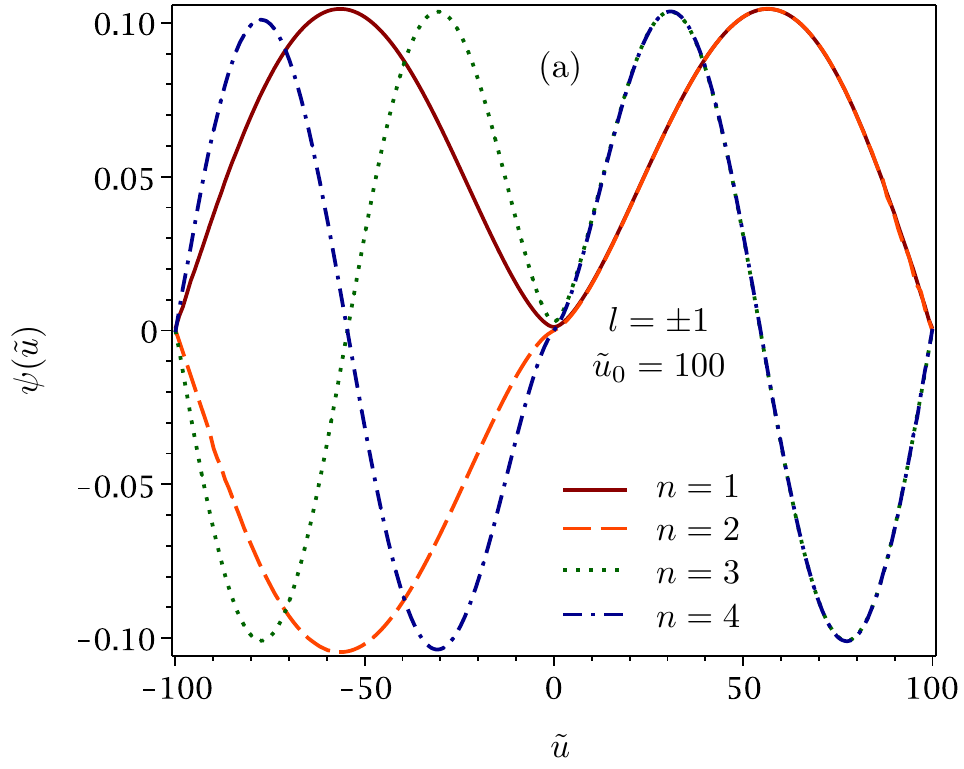} \\
        \vspace{0.3cm}
        \includegraphics[width=0.95\columnwidth]{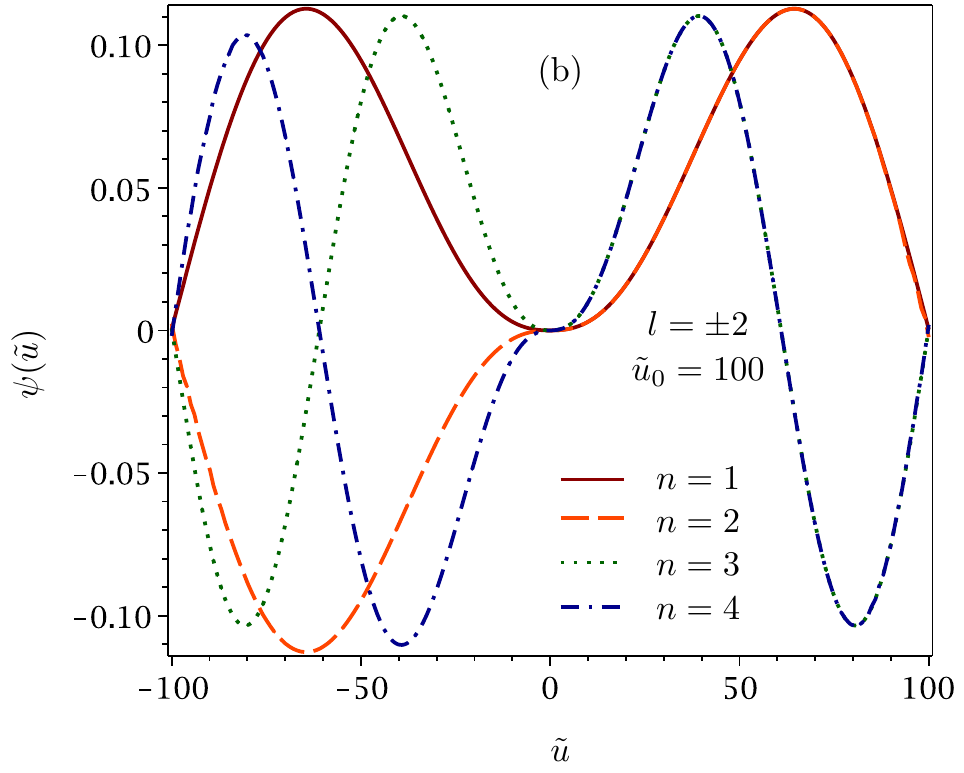}
        \caption{Numerically normalized radial wave functions  corresponding to the first four states (a) for $l = \pm 1$ and (b) for $l = \pm 2$, \textit{i.e.}, for the repulsive effective potential, and considering $\tilde{u}_{0} = 100$.}
        \label{fig:repulsive-l}
\end{figure}

When we set $\tilde{u}_{0} = 100$, the energies for the first four states where $l = \pm 1$ ($l = \pm 2$) are given approximately by $0.00146$ ($0.00264$), $0.00147$ ($0.00264$), $0.00490$ ($0.00708$), and $0.00491$ ($0.00708$). From these results, we show in Fig. \ref{fig:repulsive-l}  the corresponding wave functions. There, it is possible to observe in both $l = \pm 1$ and $l = \pm 2$ cases that the probability densities $|\psi|^{2}$ are very similar for states where $n = 1$ and $n = 2$, as well as for $n = 3$ and $n = 4$, and so on. Furthermore, the effects of the repulsive potential become evident here, since the probability density near the origin is practically null in all the studied states. Again, all of this indicates that the possible energies of the particle start to form a continuous spectrum for the repulsive effective potential when the values of $ \tilde{u}_{0}$ are large enough.

\section{On the ground state}
\label{sec:ground-state}
\begin{figure}[htp]
    \centering
    \includegraphics[width=0.8\columnwidth]{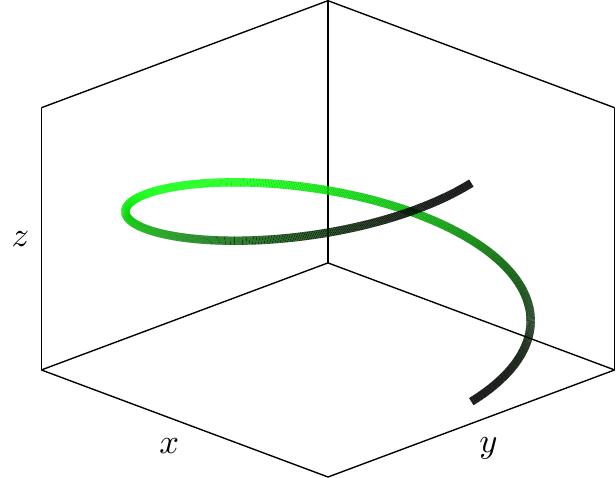}
    \caption {Color map representation of the probability density $|\psi_{1}|^{2}$ corresponding to the ground state of a particle constrained to a finite helix with $\phi_{0} = 2\pi$. The brighter (darker) the region, the greater (smaller) the probability of finding the particle.}
    \label{fig:helix-dens-prob}
\end{figure}

In subsection \ref{subsec:helix-results} we showed that, in the CPF, a quantum particle constrained to an infinite helix has its angular momentum quantized, such that the probability density is always constant, since $|\psi_{l}(\phi)|^{2} = \dfrac{1}{2\pi}$, for any integer $l$. On the other hand, if the helix is finite for $0 < \phi < \phi_{0}$, thus the wave functions $\psi_{j}(\phi)$ and the energies $E_{j}$ are analogous to the one-dimensional infinite potential well problem. Figure \ref{fig:helix-dens-prob} shows the probability density of the ground state, $|\psi_{1}|^{2}$, plotted on a finite helix with $\phi_{0} = 2\pi$.

\begin{figure}[htp]
    \centering
    \includegraphics[width=0.7\columnwidth]{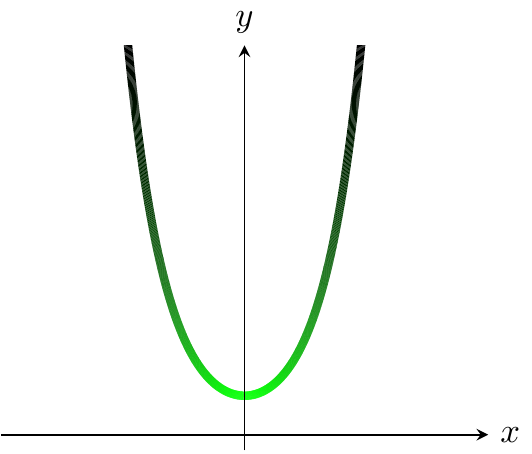}
    \caption {Color map representation of the probability density $|\psi|^{2}$ corresponding to the ground state of a particle constrained to a catenary. The brighter (darker) the region, the greater (smaller) the probability of finding the particle.}
    \label{fig:catenary-dens-prob}
\end{figure}

Furthermore, we showed in  section \ref{radial-part} that a particle constrained to a catenary can be found in a bound state due to the GIP. When we make $l = 1/2$ and $\tilde{u} = u/a$ in Eq. \eqref{sol-odd}, we get the wave function
\begin{equation}
    \begin{split}
        \psi(u) & = \dfrac{C_{0}}{\sqrt{a}}\left(1 + \dfrac{u^{2}}{a^{2}} \right)^{\frac{\sqrt{5}}{4} + \frac{1}{2}} \times \\
        & \ \times \HeunC\left(0, -\dfrac{1}{2}, \dfrac{\sqrt{5}}{2}, -\dfrac{\epsilon}{4}, \dfrac{9}{16} + \dfrac{\epsilon}{4}, -\dfrac{u^{2}}{a^{2}}\right),
    \label{catenary-ground}
    \end{split}
\end{equation}
where $C_{0} \approx 0.37842$ e $\epsilon \approx - 0.02892$, according to the box  method. Thus, the energy of the particle in the bound state is given by $E \approx - 0.02892 \dfrac{\hbar^{2}}{2ma^{2}}$. Using \eqref{catenary-ground}, the probability density plotted on the catenary is shown in Fig. \ref{fig:catenary-dens-prob}.

\begin{figure}[htp!]
        \centering     
        \includegraphics[width=0.85\columnwidth]{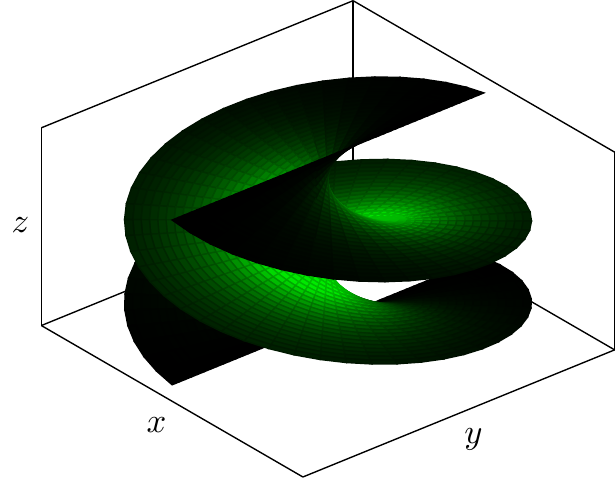}
        \caption {Color map representation of the probability density $|\chi_{g}|^{2}$ corresponding to the ground state of a particle constrained to a finite helicoid with $\phi_{0} = 2\pi$. The brighter (darker) the region, the greater (smaller) the probability of finding the particle.}
        \label{fig:helicoid-dens-prob}
 \end{figure}

For a helicoid finite  in the $\phi$ variable, with $\phi_{0} = 2\pi$, we obtain from Eq. \eqref{helicoid-finite} the angular wave function,
 \begin{equation}
     \Phi_{j}(\phi) = \dfrac{1}{\sqrt{\pi}} \sin\left(\dfrac{j\phi}{2}\right),
 \end{equation}
thus $j = 1, 2, 3,\dots$ and $l = j / 2$. Since the effective potential in \eqref{effpot} is attractive only if $|l| \leq 1/2$, thus, for the helicoid in question, this implies that it is only possible to have bound states if $j = 1$. Therefore, the radial wave function in this case is also given by Eq. \eqref{catenary-ground}, such that the total wave function of the single bound state is given by
\begin{equation}
    \begin{split}
        \chi_{g} (\phi, & u) = \dfrac{C_{0}}{\sqrt{\pi a}} \sin \left(\dfrac{\phi}{2}\right) \left(1 +  \dfrac{u^{2}}{a^{2}}\right)^{\frac{\sqrt{5}}{4} + \frac{1}{2}} \times \\
        & \ \times \HeunC\left(0, -\dfrac{1}{2}, \dfrac{\sqrt{5}}{2}, -\dfrac{\epsilon}{4}, \dfrac{9}{16} + \dfrac{\epsilon}{4}, -\dfrac{u^{2}}{a^{2}}\right),
        \label{helicoid-ground}
    \end{split}
\end{equation}
where $C_{0} \approx 0.37842$ and $\epsilon \approx - 0.02892$. Thus, the energy of the particle in the bound state is $ E \approx - 0.02892 \dfrac{\hbar^{2}}{2ma^{2}}$, as in the case of the catenary. Using Eq. \eqref{helicoid-ground}, the probability density plotted on the helicoid is shown in Fig. \ref{fig:helicoid-dens-prob}.

\begin{figure}[htp!]
        \centering 
         \hspace{-6cm} (a) \\ 
        \includegraphics[width=0.8\columnwidth]{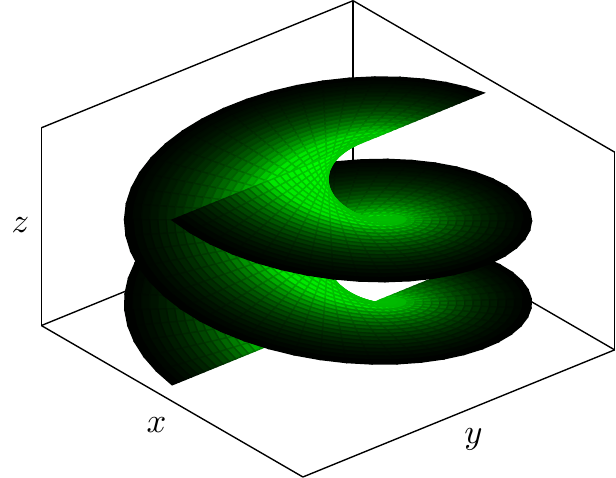} \\
         \hspace{-6cm} (b) \\ 
        \includegraphics[width=0.8\columnwidth]{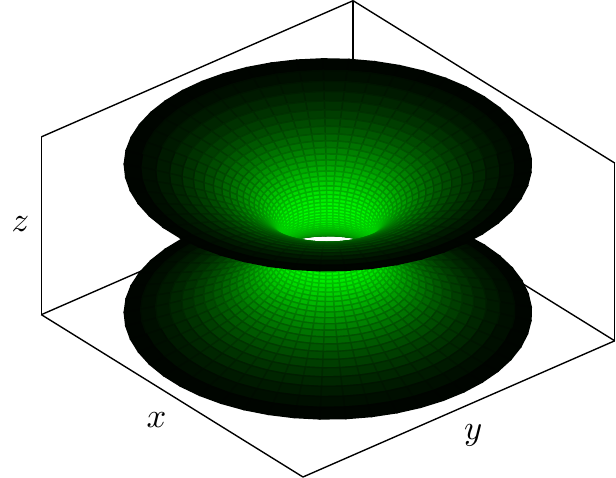}
        \caption {Color map representation of the probability density $|\chi_{g}|^{2}$ corresponding to the ground state of a particle constrained (a) to a infinite helicoid and (b) to a catenoid. The brighter (darker) the region, the greater (smaller) the probability of finding the particle.}
        \label{fig:catenoid-dens-prob}
\end{figure}

Finally, as the angular wave function for the cases of the infinite helicoid and the catenoid is given by $\Phi_{l}(\phi) = \dfrac{1}{\sqrt{2\pi}} e^{il\phi}$, and the effective potential is attractive only if $l = 0$, thus, for there to be bound states, we simply have $\Phi_{0}(\phi) = \dfrac{1}{\sqrt{2\pi}}$. From the radial wave function \eqref{sol-odd}, the total wave function for the single bound state is
\begin{equation}
    \begin{split}
        \chi_{g}(\phi, & u) = \dfrac{C_{0}}{\sqrt{2\pi a}} \left(1 + \dfrac{u^{2}}{a^{2}}\right)^{\frac{\sqrt{5}}{4} + \frac{1}{2}} \times \\
        & \ \times \HeunC\left(0, -\dfrac{1}{2}, \dfrac{\sqrt{5}}{2}, -\dfrac{\epsilon}{4}, \dfrac{5}{8} + \dfrac{\epsilon}{4}, -\dfrac{u^{2}}{a^{2}}\right),
        \label{catenoid-ground}
    \end{split}
\end{equation}
where $C_{0} \approx 0.49087$ and $\epsilon \approx - 0.13051$, as found with the box  method. Thus, the energy of the particle is given by $ E \approx - 0.13051\dfrac{\hbar^{2}}{2ma^{2}}$. Figure \ref{fig:catenoid-dens-prob} shows the probability density plotted on the respective surfaces, as obtained from Eq. \eqref{catenoid-ground}.

{ \section{Topological aspects}
\label{topological-aspects}
As mentioned before, the helicoid and catenoid are {locally isometric to each other, that is, their local geometry is indistinguishable.} But, as it is obvious from Figs. \ref{fig:helicoid} and \ref{fig:catenoid}, they are quite different from a global point of view. In fact, these surfaces have different topologies. Any closed curve on the helicoid can be shrunk to a point, while there is a class of loops on the catenoid that will end up stuck at the catenoid's neck upon reduction. Quantum mechanics, being inherently delocalized, offers many possibilities as a probe for topology. A well known  example, the Aharonov-Bohm \cite{aharonov1959significance} phase stands as a  self-evident quantum observable that is directly related to topology.} 

{ Topology has become an important tool both in the study of the physical properties as well as in the design of new materials \cite{gupta2014topological}. Physical systems may have very rich hidden topologies, like in Fermi surfaces, for instance. A change of  topology of the Fermi surface signals an electronic phase transition, as observed  by Lifshitz \cite{lifshitz1960anomalies} already in 1960. A large class of modern materials, the so-called Topological Materials \cite{vergniory2019complete}, rely their unusual, and sometimes exotic properties, in the intrinsic topology of their electronic band structures.   A detailed account of the  role of topology in modern materials is given in Ref. \cite{gupta2018role}. It is clear then, the importance of probing the topology of a physical system by means of  specific measurements. The concept of using the measurement of a physical observable to identify the  topology of a  system, or topological metrology \cite{gupta2014topological}, becomes then an important tool for the geometric/topological characterization of novel materials. How to apply this idea to the systems studied here? How an experiment can distinguish between two isometric surfaces with different topologies? A hint comes from the aforementioned Aharonov-Bohm phase. The original A-B setup \cite{aharonov1959significance} involves a non simply-connected (shielded) region of space which is threaded by a magnetic flux.  Electron beams traveling on either side of the ``hole'' that contains the flux acquire different quantum phases whose difference can be measured by interference. This establishes the A-B phase as a topological metric in the sense described above. On the other hand, geometry and topology give rise to analog A-B effects. For instance, the parallel transport of vectors and spinors around cosmic strings (literally a cylindrical hole in spacetime)  gives rise \cite{ford1981gravitational} to nontrivial phases that  reflects the locally flat spacetime geometry of this cosmic object and its conical topology. Disclinations, which are topological line defects in nematic liquid crystals, present a similar effect for light propagation \cite{carvalho2007aharonov}. A measurement of the phase acquired by the polarization (a spinor) of the light traveling around the disclination can give  information on its topology. Both the cosmic string and disclination examples can be unified under the single view of a A-B-type effect where the magnetic field is substituted by a curvature flux \cite{carvalho2013holonomy}. 
}

\begin{figure}[ht]
    \centering
 \includegraphics[width=0.8\columnwidth]{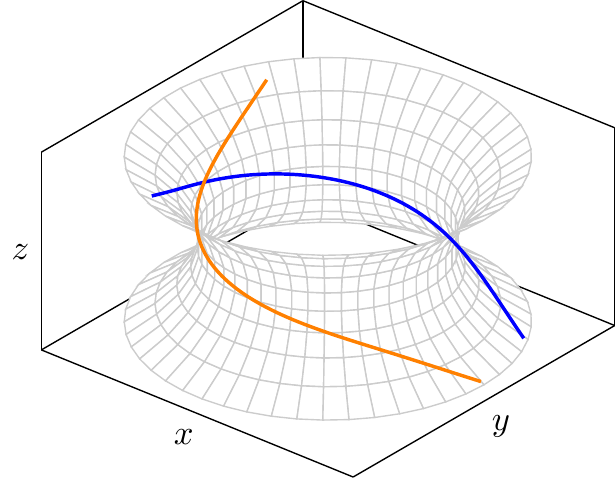}
    \caption{{ Representation of  equivalent classical trajectories of electrons injected at the edge of the catenoid.}}
    \label{fig:geodesics}
\end{figure}

{Wavefunctions, being scalar, are trivially parallel-transported, not yielding any topological information in this way. Nevertheless, upon scattering, the acquired phase shift of the wave function might provide topological information as in the case of diffraction of light by disclinations in liquid crystals \cite{pereira2011diffraction}. The phase shift appears then as possible topological metric. Another option comes from higher order tensors like vectors or spinors which, as discussed above, may reveal topological features when subjected to parallel transport. In modeling curved graphene with the Dirac equation, one naturally deals with spinors, which can then {be} used as probes for topology in the A-B scheme. Recent tight-binding results \cite{stegmann2018current} indicated that curvature of elastically deformed graphene can be used to split the electric current into two beams of differently valley-polarized electrons. The curvature, represented by the strain field,   acts on the electrons as a pseudo-magnetic field splitting the otherwise degenerate valley states. This could be a strategy for an Aharonov-Bohm-type of experiment that would compare the phases acquired by the valley spinors in each beam as they travel on opposing sides of the curved region (see Fig. 1 of Ref. \cite{stegmann2018current}). Then, parallel transport of the spinors, measured by interference of valley-polarized electron beams traveling along equivalent paths, like the ones shown on the catenoid of Fig. \ref{fig:geodesics},  can be used as a tool for topological metrology and thus distinguish between two isometric surfaces with different topologies, like the ones studied here. }

\section{Conclusions}
\label{conclusions}
{In this work,  we studied the confinement of a quantum particle to a helix, catenary, helicoid, or catenoid using the  Confining Potential Formalism to curves and surfaces. } First, it was shown that a particle constrained to an infinite helix has its angular momentum quantized. On the other hand, for the confinement to a finite helix, the problem becomes similar to that of a quantum particle in a one-dimensional potential well.

The local isometry between the finite helicoid and the catenoid was explored, such that it was shown that the quantum dynamics on the surfaces is governed by the same Schrödinger equation. From this, an effective Schrödinger equation was obtained, in the radial variable $\tilde{u}$, with an effective potential $V_ {l}(\tilde{u})$ that is attractive only if $|l|\leq 1/2$. In the case of the infinite helicoid and the catenoid, it is only possible to have bound states when $l = 0$, while in the case of a  helicoid limited to $0 < \phi < 2 \pi $ this occurs only for $l = 1/2$. In addition, it was observed that, by making $l = 1/2 $ in the radial equation for the helicoid and the catenoid, the same Schrödinger equation is obtained for the case of the catenary.

The use of the variational method (with the Gaussian and the Lorentzian functions) guarantees the existence of a bound state in the cases of the catenary, the helicoid (finite and infinite in $\phi$) and the catenoid. Thus, it was shown that the radial Schrödinger equation has exact solutions that are written in terms of the confluent Heun function, however, the eigenvalues cannot be obtained analytically, so it was necessary to use numerical methods to estimate them. In fact, both methods were in agreement with each other, and it was possible to infer that, in these three cases, the particle can be found in a single bound state, which is due to the GIP that comes from the CPF.

To sum up, we recall that Refs. \cite{atanasov2009geometry,dandoloff2010geometry,silva2020electronic}  provided detailed analysis of the geometry-induced quantum potential and some of its effects due to confinement of a particle to either a helix, catenary, helicoid, or catenoid. We  extend and complement these previous works by providing a unified treatment of the Schr\"odinger equations of the corresponding cases. Furthermore, we determine the single bound state that appears in the infinite versions of the catenary, helicoid, and catenoid and show that the remaining states combine into a continuous energy band.

As perspectives for future works, we intend to study the confinement of a quantum particle with spin in the helicoid and catenoid, as well as to study the confinement of identical particles under these conditions. Since the spin is sensitive to  torsion \cite{dandoloff2004quantum}, it may be able to distinguish the surfaces in spite of their isometry. Another interesting approach is the addition of a geometry-induced magnetic field to the helicoid as done in Ref. \cite{wang2020geometry}. We also intend to use the methods used in the present work for the confinement problem in other regular curves and surfaces.

\acknowledgements
This study was financed in part by the Coordenação de Aperfeiçoamento de Pessoal de Nível Superior – Brasil (CAPES) – Finance Code 001 (F.F.S.F.),  Fundação de Amparo à Ciência e Tecnologia do Estado de Pernambuco (FACEPE), Grants No. IBPG-0487-1.05/19 (J.D.M.L.), and BIC-1187-1.05/20 (E.G.), and Conselho Nacional de Desenvolvimento Científico e Tecnológico (CNPq), Grant No. 307687/2017-1 (F.M.). The authors are indebted to L.C.B. da Silva for invaluable discussions and suggestions.

\providecommand{\noopsort}[1]{}\providecommand{\singleletter}[1]{#1}%

\end{document}